\documentclass[11pt,a4paper]{article}
\pdfoutput=1

\usepackage{hyperref}
\usepackage{bm}
\usepackage{amsmath,amssymb}
\usepackage{cite}

\usepackage{graphicx}

\numberwithin{figure}{section}
\numberwithin{equation}{section}

\newcommand{\be}{\begin{equation}}
\newcommand{\ee}{\end{equation}}
\newcommand{\bea}{\begin{eqnarray}}
\newcommand{\eea}{\end{eqnarray}}

\newcommand{\vp}{\varphi}
\newcommand{\pb}{\bar \varphi}

\newcommand{\al}{\alpha}
\newcommand{\Vh}{\hat V}
\newcommand{\ph}{\hat \varphi}
\newcommand{\that}{\hat t}
\newcommand{\vh}{\hat v}

\newcommand\ie{\textit{i.e.}\ }
\newcommand\eg{\textit{e.g.}\ }
\newcommand\cf{\textit{cf.}\ }
\newcommand{\eps}{\varepsilon}

\newcommand\tc[1]{\raisebox{.5pt}{\textcircled{\raisebox{-.9pt} {#1}}}}

\begin{document}

\begin{titlepage}

\begin{center}
{\huge \bf  The local potential approximation in the 
\vspace{0.3cm}

background field formalism}
\end{center}
\vskip1cm


\begin{center}
{\bf I. Hamzaan Bridle, Juergen A. Dietz and Tim R. Morris}
\end{center}

\begin{center}
{\it School of Physics and Astronomy,  University of Southampton\\
Highfield, Southampton, SO17 1BJ, U.K.}\\
\vspace*{0.3cm}
{\tt  ihb1g12@soton.ac.uk, J.A.Dietz@soton.ac.uk,\\ T.R.Morris@soton.ac.uk}
\end{center}

\abstract{Working within the familiar local potential approximation, and concentrating on the example of a single scalar field in three dimensions,
we show that the commonly used approximation method of identifying the total and background fields,
leads to pathologies in the resulting fixed point structure and the associated spaces
of eigenoperators. We then show how a consistent treatment of the background field through
the corresponding modified shift Ward identity, can cure these pathologies,  restoring universality of physical quantities with respect to the choice of dependence on the background field,
even within the local potential approximation.
Along the way we point out similarities to what has been previously found in the $f(R)$ approximation
in asymptotic safety for gravity.}



\end{titlepage}

\tableofcontents

\section{Introduction}
\label{sec:Intro}
The functional renormalisation group formulation of Wilson's  renormalisation group \cite{Wilson:1973}
has been employed
in a wide variety of contexts and has proved a powerful tool in many instances (see \eg \cite{Berges:2000} for a review).
The philosophy behind this technique is that it can prove useful for practical purposes 
to integrate out only those modes in the path integral that possess momenta larger than some infrared 
cutoff scale $k$. Varying $k$, a non-perturbative renormalisation group flow is generated with
the property that
 upon sending $k \rightarrow 0$ we recover the information contained in the full
path integral \cite{Morris:1998,Bagnuls:2000,Rosten:2010}.

The essential ingredient in this approach is the cutoff action $S_k$ which is added 
to the bare action leading to the desired modification of the path integral.
For the purposes of the discussion here it suffices to phrase things in terms of a single scalar field $\phi$.
Commonly the cutoff action is then taken to be of the form
\begin{equation} \label{equ:cutoffaction}
 S_k[\phi] = \frac{1}{2} \int_x \phi\, R_k\!\left(-\nabla^2\right)\phi,
\end{equation}
where we have abbreviated integration over space by the corresponding subscript. The cutoff operator
$R_k\left(-\nabla^2\right)$ 
 turns $S_k$ into a momentum dependent mass term by acting on the field $\phi$.
In order to obtain well defined renormalisation group flows, the cutoff operator has to satisfy some constraints
but to a large extent can be chosen arbitrarily 
 (see \eg \cite{Litim:2001}).

In principle, physical results obtained from the renormalisation group flow are independent of the 
choice of cutoff operator. In practice however, in order to make actual computations possible, one has to
resort to various approximations in setting up the flow, with the effect that physical quantities 
depend on the cutoff implemented.\footnote{Actually, even at the level of an exact treatment, apparently sensible choices of cutoff can potentially lead to
unphysical behaviour or even cause the breakdown of the flow itself \cite{Hasenfratz:1987,Sokal:1994}.} We will be interested in one such very common approximation, which arises in the context of Yang-Mills theory or gravity. In these cases, the background field method is nearly always employed since it ensures that the results of computations are expressed covariantly (see \eg
\cite{Reuter:2012,Reuter:1993kw}). It does so however at the expense of introducing two fields: the original (total) field $\phi$ and a separate background field $\pb$, and at the expense of introducing an in-principle arbitrary dependence of the cutoff operator on the background field (apart from required covariance). 

The renormalisation group flows considered in this work are formulated in terms of the so-called effective
average action \cite{Wetterich:1992,Berges:2000,Reuter:1993kw,Gies:2006wv,Pawlowski:2005xe},
which in the context of the background field method is a functional of both fields.
As we review in sec. \ref{sec:LPAsetup}, this problem is commonly dealt with by neglecting the part of the effective average action that captures the deviation from $\phi^c=\pb$, where $\phi^c$ is the classical
counterpart of the quantum field $\phi$, allowing the two fields to be identified from then on.
As a result, the effective average action is again a functional of only one field. 
We will see however that this single field approximation
can readily lead to severe pathologies, and we will stress the relevance for the $f(R)$ approximation in the asymptotic safety programme for quantum gravity. We will then show that there is an elegant way out, at least within the scalar field theory model we deal with here, which allows one to recover exactly the desired universality of physical quantities with respect to rather general choices of dependence of the cutoff on the background field.

In the scalar field theory model that we will use, our strategy will thus be to allow the cutoff operator $R_k$ to be a function(al) of momentum and the field 
itself. Following the standard route, this requires the use of the background field method, as the cutoff operator can 
only depend on the background field and not on the total field, if we are to keep the familiar form of the flow equation for the effective average action.
In terms of the cutoff action \eqref{equ:cutoffaction} this means that we generalise the form of the cutoff operator to
$R_k\left(-\nabla^2,\pb\right)$. 
 As we have already indicated, we will then adopt the single field approximation which in effect identifies the total and background fields. 
We choose the field dependence of $R_k$ to be 
controlled by an additional parameter $\al$ such that for $\al=0$ the dependence on the field drops out.
We can then dial this parameter $\al$ and investigate the effects on the well known results for the fixed point
structure and critical exponents of scalar field theory in the Local Potential Approximation (LPA) \cite{NCS,Hasenfratz:1985dm,Morris:1994ki}, where for most of what follows we will restrict ourselves to a single scalar field in three Euclidean dimensions\cite{Morris:1994ki,Morris:1994ie, Bervillier:2007, Litim:2002cf}.
In sec. \ref{sec:LPA} we will investigate the resulting renormalisation group flows for two very different choices of background field dependence of the cutoff operator. The first choice (sec. \ref{sec:gencutoff}) is an explicit background field dependence, whereas for the second choice (sec. \ref{sec:LPALitim}) the background field dependence
comes in through the effective average action itself. We then find in both cases that 
 as soon as the cutoff operator becomes field dependent
we encounter severe deviations from the standard picture for both the fixed point structure
as well as the spaces of eigenoperators. These deviations are such that the physical content
characterising the Gaussian and Wilson-Fisher fixed points for three dimensional scalar field theory is only partly preserved or even completely absent, 
depending on the value of $\al$.

Nevertheless, as described in sec. \ref{sec:backgroundfield}, there is a way to overcome these
problems which requires a proper treatment of the background field. This is possible through
what we call the modified shift Ward identity which takes into account the field dependence of the cutoff operator
and complements the renormalisation group flow as defined by the flow equation. 
 We show explicitly how the physical properties of the fixed points for scalar field theory in three dimensions  can be recovered both qualitatively and quantitatively at the level of the LPA, for a large class of background field dependent cutoff operators $R_k$
which includes the two choices explicitly investigated in sec. \ref{sec:LPA}.
In other words, for scalar field theory in the LPA the modified shift Ward identity makes it possible
to restore universality with respect to rather general $k$ and field dependence in $R_k$.

These results have to a certain extent been anticipated in the literature, in particular the modified shift Ward identity has been discussed extensively. However it is most often expressed as a formula for the differential dependence of the effective average action on changes to the background field $\pb$ taken at constant total classical field $\phi^c$. For scalar field theory it appears in this form in \cite{Litim:2002hj},  and for gauge theories in refs. \cite{Reuter:1993kw,Litim:1998nf,Litim:2002ce,Reuter:1994sg,Reuter:1997gx}.   In refs. \cite{Manrique:2009uh,Manrique:2010mq,Manrique:2010am}  they  refer to it as the violation of ``split symmetry'', and it is recognised as having a significant impact on the fixed point structure, in the asymptotic safety programme in quantum gravity in particular \cite{Weinberg:1980,Reuter:1996,Reuter:2012}. 

What is new to this paper is however on the one hand the much fuller recognition of the extent and severity of the possible pathologies involved if the modified Ward identity is not respected and on the other hand the discovery that in scalar field theory the simple expedient of modifying the LPA by substituting the modified shift Ward identity is actually sufficient to recover exact universality. 
In addition, we emphasise in sec. \ref{sec:backgroundfield} that the flow equation on its own is not enough 
to uniquely determine the form of the effective average action when the background field method is used without the
single field approximation. Instead, the modified shift Ward identity has to be imposed as an additional constraint.

Our study is primarily
 motivated by the use of this framework  to look for ultraviolet fixed
points in the theory space of quantum gravity \cite{Weinberg:1980,Reuter:1996,Reuter:2012} and in particular
some very recent results \cite{DietzMorris:2013-2}.
In the study of the so-called $f(R)$ approximation to quantum gravity in which the action
is truncated to a general function $f(R)$ of the Ricci scalar $R$, it was found in ref.\cite{DietzMorris:2013-2}
that, in one particular formulation of the truncated flow \cite{Benedetti:2012}, all eigenoperators become redundant,
\ie there is nothing left to explore in the sense that the space of (essential) eigenoperators is empty. At first sight it may seem that this is just a peculiarity associated to the type of truncation
considered. However, by mimicking at least in part
the approach employed in the derivation of the flow equation 
in the $f(R)$ truncation for gravity in the familiar context of the LPA for scalar field theory, we find that very similar problems appear, suggesting that the current approach adopted in the derivation
of flow equations in asymptotic safety for gravity may 
 have to be revised, as we discuss in the concluding section.

\section{LPA with field dependent cutoff operators}\label{sec:LPA}
\subsection{Setup and derivation of flow equation} \label{sec:LPAsetup}
The key ingredient for our investigations is the functional renormalisation group equation \cite{Wetterich:1992,Morris:1993}
\begin{equation} \label{equ:FRGE}
\frac{\partial}{\partial t}\Gamma_k = \frac{1}{2} \, \mathrm{Tr} \left[ \left(\Gamma^{(2)}_k + R_k\right)^{-1} \frac{\partial}{\partial t} R_k \right],
\end{equation}
formulated in terms of the scale dependent effective average action $\Gamma_k$. In this expression, $\Gamma_k^{(2)}$ denotes
the second functional derivative with respect to the fields, $R_k$ is the cutoff operator from \eqref{equ:cutoffaction}
and $t=\ln(k/k_0)$ denotes  the so-called renormalisation group time, $k_0$ being a fixed physical reference scale.
The trace runs over all indices of the field as well as over space (or momentum) coordinates. All calculations
are carried out in Euclidean signature.

Our goal will be to allow the cutoff operator $R_k$ to depend on the field itself which implies 
that the cutoff action \eqref{equ:cutoffaction} is no longer quadratic in $\phi$. In general, such a modification
would spoil the appealing one-loop structure of the flow equation \eqref{equ:FRGE} whose derivation depends on
the cutoff action $S_k$ being quadratic in the fields. This obstacle can be circumvented by applying the background field
method. In this approach we split the bare quantum field $\phi$ into a fixed background field $\pb$ and a fluctuation field
$\vp$ according to
\begin{equation*}
 \phi = \pb + \vp
\end{equation*}
and it is the fluctuation field $\vp$ which is sourced in the path integral. Using standard techniques
(\eg \cite{Wetterich:1992}) it is then straightforward to show that the flow equation \eqref{equ:FRGE} still holds
provided the cutoff action \eqref{equ:cutoffaction} is quadratic in the fluctuation field and thus $R_k=R_k\!\left(-\nabla^2,\pb \right)$. 
The effective average action is now a functional of both the classical field $\vp^c$ and the background field,
$\Gamma_k = \Gamma_k[\vp^c,\pb]$, and the Hessian on the right hand side of the flow equation is taken with respect to the
classical field.

Employing the background field method in this way to obtain greater flexibility in the choice of cutoff
comes at the price of having to deal with an effective average action depending on two arguments: the classical and
the background field. In common with most of the literature on functional renormalisation group methods in 
asymptotic safety for gravity where one faces the same complication,
we are going to adopt the following approximation.
Let us decompose\footnote{\label{Reuter-vars} This discussion follows ref. \cite{Reuter:1996}, where it is expressed in terms of ${\tilde\Gamma}_k[\phi^c,\pb] \equiv \Gamma_k[\phi^c-\pb,\pb]$} 
\begin{equation} \label{approx}
 \Gamma_k[\vp^c,\pb]= 
 \Gamma_k[\phi^c]+ \hat \Gamma_k[\vp^c,\pb],
\end{equation}
where we have defined $\phi^c=\pb+\vp^c$ as the classical counterpart of the full quantum field, and $\Gamma_k[\phi^c] = \Gamma_k[0,\phi^c]$ is defined by substituting the total classical field for the background field.
By definition therefore, the remainder vanishes for vanishing (classical) fluctuation field: $\hat \Gamma_k[0,\pb]=0$.
In this sense, $\hat \Gamma_k$ captures the deviation in the effective average action from $\phi^c = \pb$. 
The approximation we are going to make is to entirely neglect this second term $\hat \Gamma_k$ in the following,
setting it to zero from the outset. Terms contained in $\hat \Gamma_k$ are therefore not taken into
account in the Hessian on the right hand side of the flow equation \eqref{equ:FRGE}, nor are the corresponding
running couplings included on the left hand side. 
After the Hessian of $\Gamma_k[0,\phi^c]$ in \eqref{equ:FRGE} has been evaluated we are allowed to set $\vp^c=0$
in all subsequent calculations, thereby identifying the total classical field and the background field.
We stress that our motivation for adopting this approximation arises entirely from our aim
to follow the approach adopted in most of the asymptotic safety literature and in particular
in \cite{Benedetti:2012}.\footnote{See however refs. \cite{Lauscher:2001ya, Lauscher:2002sq, Manrique:2009uh,Manrique:2010mq,Manrique:2010am} for exploratory work which keeps some of the running $\hat \Gamma_k$, and ref. \cite{Dona:2013qba} which effectively exploits  differences between the wavefunction renormalisation of the background and fluctuation fields.}.
The advantage of making this \emph{single field approximation}  is that we now have to
deal with an effective average action depending on 
 only one field, $\phi^c$.

With this in mind we now take the effective average action to be of form $\Gamma_k[\phi^c]$.
The LPA in scalar field theory is characterised by truncating it to
\begin{equation} \label{equ:LPA}
 \Gamma_k[\phi^c] = \int_x \left( \frac{1}{2} \left(\partial_\mu \phi^c \right)^2 + V(\phi^c) \right),
\end{equation}
discarding all higher derivative terms and setting the wavefunction renormalisation to one 
(\cf \cite{Bervillier:2013}, however).
We retain a general potential $V(\phi^c)$, in particular it need not be symmetric under $\phi^c \mapsto -\phi^c$.
 For simplicity of notation, the $k$ dependence is left implicit.  
 
Let us now comment on the choice of cutoff which is the crucial feature in our study.
We choose to work with the following cutoff operator, a generalisation of Litim's cutoff \cite{Litim:2001}:
\be 
\label{equ:cutoffh}
R_k\!\left(-\partial^2,\pb(x)\right) = \left(k^2+\partial^2-h(\pb)\right)\theta\!\left(k^2+\partial^2-h(\pb)\right)\,.
\ee
Here $h$ is some general function of the background field $\pb(x)$ which may or may not itself depend on the action \eqref{equ:LPA} through $V$. 
In order to derive the flow equation originating from this form of cutoff we note that
within the LPA the field can be treated as effectively spacetime independent  \cite{Morris:1994ie}
and thus in momentum space the cutoff operator is diagonal:
\begin{equation*}
 R_k\!\left(p^2,\pb\right) = \left(k^2 - p^2 -h(\pb)\right) \theta\! \left(k^2 - p^2 -h(\pb)\right)\,.
\end{equation*}
On the right hand side of \eqref{equ:FRGE} we need the time derivative of this, given by
\begin{equation}
 \label{equ:cutoffh-tder}
 \partial_t R_k = \left(2k^2 -\partial_t h(\pb)\right) \theta \left(k^2 - p^2 - h(\pb)\right).
\end{equation}
Combining this with the second functional derivative of \eqref{equ:LPA}, the evaluation of \eqref{equ:FRGE}
leads to
\begin{equation}
 \label{equ:flowLPA}
 \partial_t V - \frac{1}{2}(d-2) \phi V' +dV = 
 \frac{(1-h)^{d/2}}{1-h+V''}\left(1-h-\frac{1}{2}\partial_t h +\frac{1}{4}(d-2)\phi h'\right)\theta(1-h),
\end{equation}
where $d$ is the space dimension.
In the process we have absorbed a constant by a field redefinition
and adopted scaled variables according to
\begin{equation} \label{scaledvars}
 \tilde \phi = k^{\frac{1}{2}(2-d)} \phi^c, \qquad \tilde V(\tilde \phi) =k^{-d}V(\tilde \phi)
  \qquad \text{and} \qquad \tilde h(\tilde \phi)=k^{-2}h(\tilde \phi),
\end{equation}
followed by an immediate renaming by dropping the tilde.

We remember that in the single field approximation we can set $\vp^c=0$ in $\phi^c=\vp^c+\pb$
once the Hessian of the ansatz \eqref{equ:LPA} has been taken. Consequently, the 
only field variable appearing in the flow equation \eqref{equ:flowLPA} is the scaled (dimensionless)
total classical field $\phi$. In particular, since we no longer distinguish between the background and
total classical fields, all appearances of $h$ in the flow equation
\eqref{equ:flowLPA} have the dependence $h=h(\phi)$.

We will study the consequences of a background field dependent cutoff operator under adoption
of the single field approximation for two qualitatively different choices of $h$. 
As our first choice, we specialise to the case where 
\begin{equation} \label{cutoff1}
h=\alpha k^{4-d}\pb^2,
\end{equation}
where $\al$ is a dimensionless free parameter introduced by hand which controls the field dependence of
the cutoff operator. 
This leads to a cutoff operator with a purely explicit dependence on the background field and we analyse this
case in sec. \ref{sec:gencutoff}.
Our second choice will give rise to a cutoff operator with an implicit background field
dependence through the potential by setting 
\begin{equation}\label{cutoff2}
h=\al V''(\pb),
\end{equation}
the dimensionless parameter $\al$ playing the same role
as in the previous example. Our investigation pertaining to this
choice is presented in sec. \ref{sec:LPALitim}.

It has previously been noted (see \eg \cite{Gies:2002af}) that the single field approximation can impose a new constraint on the regulator 
in order to obtain the correct one-loop $\beta$-function. For the ``spectral flow'' in gauge theory considered in ref. \cite{Gies:2002af} 
the constraint is $R_k(\Gamma^{(2)}_k)\to k^2$ when $\Gamma^{(2)}_k$  is sent to zero.  Without the single field approximation, after ensuring
canonical normalisation of the kinetic terms and couplings and relevant Ward identities, the only constraints on the cutoff 
functions should be those imposed by convergence (\ie that they actually do regularise the Feynman integrals). We see therefore that this extra constraint was already an albeit small signal of unphysical effects induced by the single field approximation. We reproduce spectral regularisation in our context with \eqref{cutoff2} and $\alpha=1$. In sec. \ref{sec:LPALitim} we will see that it results in severe deviations from the known results at the non-perturbative level. 

To investigate the  consequences for the scalar field theory perturbative one-loop $\beta$-function, of introducing \eqref{cutoff1} or \eqref{cutoff2} together with the single field approximation, we briefly consider eqn. \eqref{equ:FRGE} in $d=4$ dimensions. 

First consider the standard case where $h=0$. $V=V_*=1/4$ is the Gaussian fixed point. Setting 
\be
\label{perturbative}
V={1\over4}+{\lambda(t)\over4!} \,\phi^4\,,
\ee
the flow induces the usual tadpole mass term of $O(\lambda)$, as well as $\phi^{6}$ and higher power terms all of which have $O(\lambda^3)$ or higher. The only term that contributes at lowest order to the $\beta$-function  comes from the $\phi^4$ term generated by  expanding $1/(1+V'')$ to second order on the right hand side of \eqref{equ:FRGE}, where $V$ is given by \eqref{perturbative}. This yields 
\be
\label{beta-function}
\beta(\lambda) \equiv \partial_t\lambda= 6\lambda^2\,.
\ee 
Reinstating the constant $1/2(4\pi)^2$ on the RHS which was absorbed by field redefinition in \eqref{equ:FRGE}, we see that this is the correct answer as expected.

If we now take $h$ of the form \eqref{cutoff1}, as soon as $\alpha\ne0$ there is no simple solution corresponding to $V_*=1/4$. We cannot recover perturbation theory in a simple way. 

On the other hand if we take $h$ of the form \eqref{cutoff2} then the Gaussian fixed point remains as $V=V_*=1/4$. It is straightforward to confirm that the one-loop $\beta$-function again comes from expanding the right hand side of \eqref{equ:FRGE} using \eqref{perturbative} and that the $\alpha$ terms all cancel out so that one again finds the correct $\beta$-function \eqref{beta-function}. If there is a restriction analogous to that in ref. \cite{Gies:2002af} on the form of the cutoff function in order to ensure this, then it must already be implemented in this case for any $\alpha$ by our general form \eqref{equ:cutoffh}. Satisfying as this result is, it is unfortunately not enough to save this scheme from qualitatively unphysical behaviour, as we will confirm in sec. \ref{sec:LPALitim}.

In the context of the $f(R)$ truncation in asymptotic safety leading to a collapse of the space
of eigenoperators, the cutoff operators also depend on the Laplacian shifted by a field 
dependent quantity \cite{DietzMorris:2013-2,Benedetti:2012}. Although the situation there
is much more involved in the sense that the field dependence of the
cutoff operators is a mixture of the two choices \eqref{cutoff1} and \eqref{cutoff2},
in a way that moreover depends on the implementation (compare  refs. \cite{Benedetti:2012} with \cite{Codello:2008,Machado:2007}),
we expect that they again capture the essence of implementing field dependent cutoffs
and exemplify the consequences.

\subsection{The Wilson-Fisher and Gaussian fixed points.} \label{sec:WF}
In this section we first recover well-known results about three-dimensional single component scalar field theory  by setting $h=0$ in \eqref{equ:flowLPA}, in order to then investigate what are the effects of using a generalised cutoff operator
of type \eqref{cutoff1} or \eqref{cutoff2}
with $\al \neq 0$ on both the fixed point structure and the eigenspectra in the following sections. Correspondingly, we
will set $d=3$ from now on.

For $h=0$ the cutoff operator \eqref{equ:cutoffh} reduces to the standard version
of the optimised cutoff \cite{Litim:2001} which has been used before in the context of three dimensional scalar field theory. We are here going to reproduce results that can be found for example in
\cite{Litim:2002cf} but we nevertheless go through their derivation as our approach differs from
the one used there and we will 
follow the same steps for the more complicated flow equation \eqref{equ:flowLPA} when $h \neq0$
in the next sections.

Eliminating the field dependence in the cutoff operator by setting $h=0$ in \eqref{equ:flowLPA} and focussing on scale invariant 
fixed point potentials, we obtain the fixed point equation
\begin{equation}
 \label{equ:fpWF}
 3V_*- \frac{1}{2} \phi V'_* = \frac{1}{1+V''_*}.
\end{equation}

We are first interested in the behaviour of the potential at large field values. Assuming that at zero order we 
can neglect the quantum corrections on the right hand side of \eqref{equ:fpWF} we find 
$V_0(\phi)=A\phi^6$ as the solution of the left hand side, $A$ being a real constant.
Of course, substituting this into the right hand side, we see that for $A\ne0$ the corrections are indeed subleading.
At first order, we use $V_*(\phi)=V_0(\phi)+V_1(\phi)$ on the left hand side of \eqref{equ:fpWF}
and equate this to the quantum corrections stemming from $V_0$ only.
At this order we keep only the leading term in a Taylor expansion in $1/\phi$ of the right hand side 
which leads to
\begin{equation}
 \label{equ:Vasy1}
 3V_1- \frac{1}{2} \phi V_1' = \frac{1}{30A\phi^4}\,.
\end{equation}
The general solution of this equation just reproduces the leading term $V_0$ but the special solution
supplies us with the subleading contribution $V_1(\phi)=1/150A\phi^4$.
Continuing with this process the asymptotic series for large $\phi$ can be developed
to arbitrary order, the first terms of which are:
\begin{equation}  \label{equ:VWFasy}
V_*(\phi) = A{\phi}^{6}+{\frac {1}{150}}\,{\frac {1}{A{\phi}^{4}}}-{\frac {1}{6300
}}\,{\frac {1}{{A}^{2}{\phi}^{8}}}+{\frac {1}{243000}}\,{\frac {1}{{A}
^{3}{\phi}^{12}}}-{\frac {1}{67500}}\,{\frac {1}{{A}^{3}{\phi}^{14}}} + \dots 
 \end{equation}
We note that this series depends on only one free parameter $A$ and, after mapping onto
the corresponding variables, the first two terms of this expansion coincide with ref. \cite{Litim:2002cf}.

\begin{figure}[ht]
  \begin{center}
  $
   \begin{array}{cccc}
     \includegraphics[width=0.45\textwidth,height=0.295\textheight]{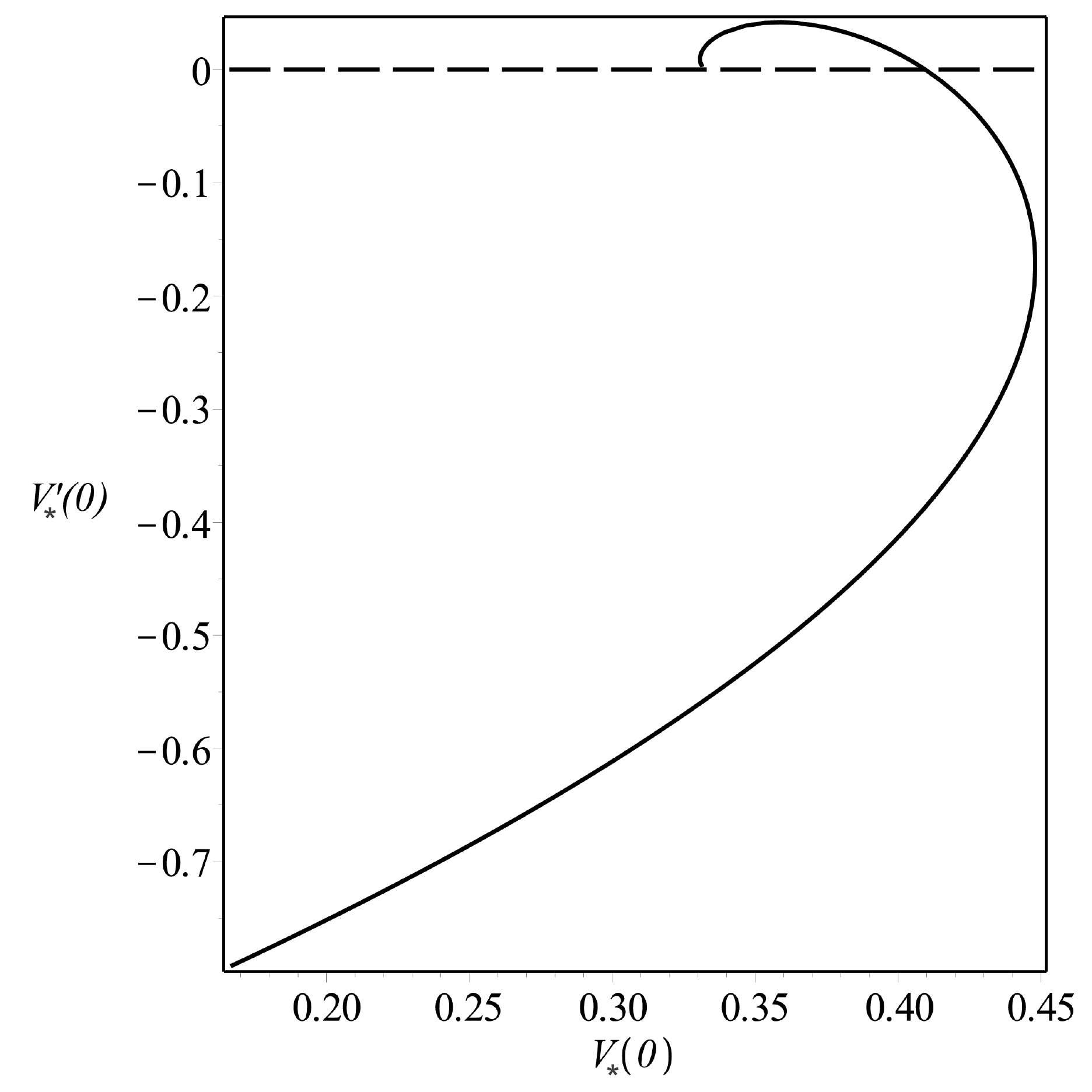} & \hfill &  \hfill &
     \includegraphics[width=0.4\textwidth,height=0.3\textheight]{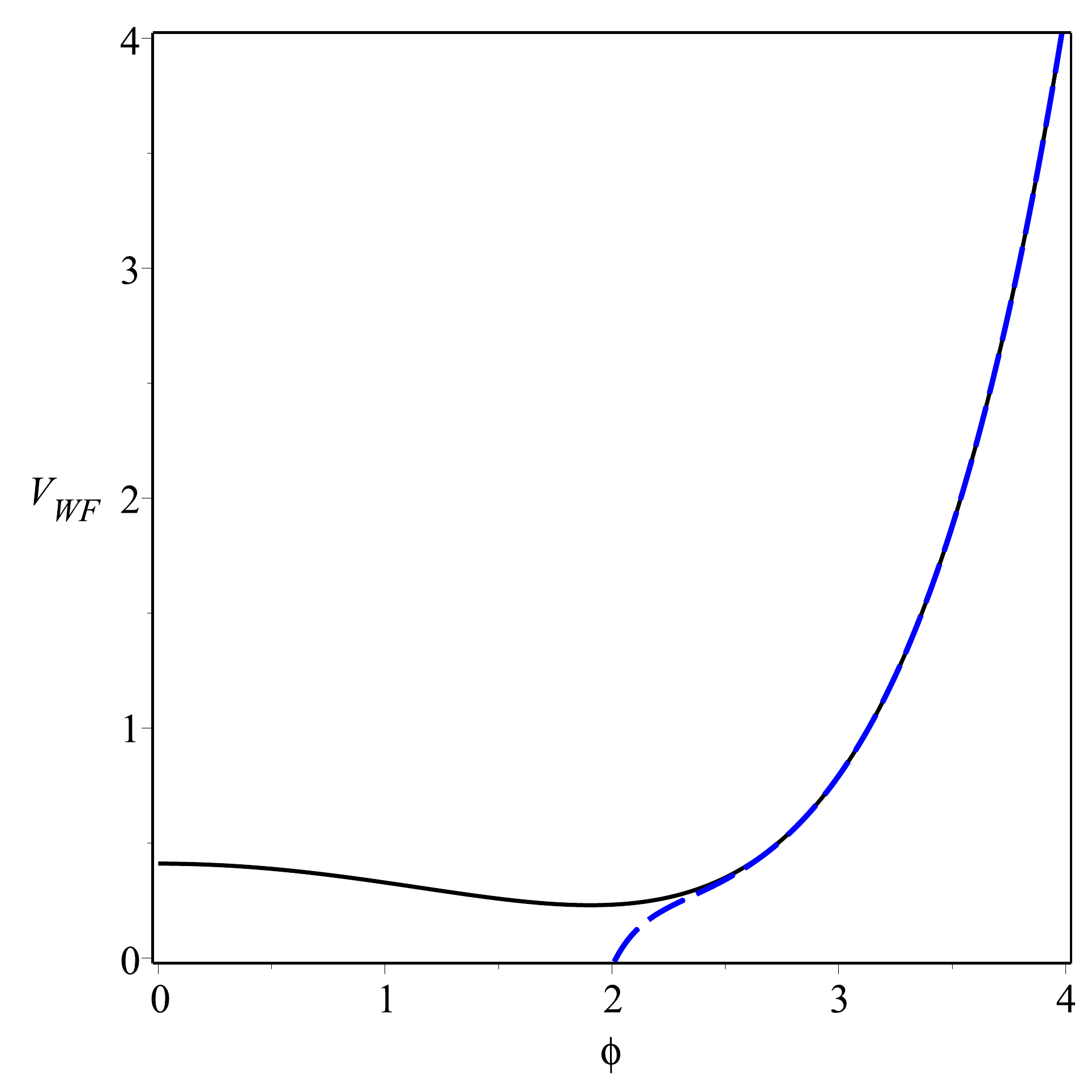} 
   \end{array}$
  \end{center}
   \caption{Varying the asymptotic parameter $A$ from $A\to0$ at the top to $A=50$ at the bottom we plot
	$V'_*(0)$ against $V_*(0)$ obtained by numerical integration (left). The Wilson-Fisher fixed point
	potential in black and solid (right) together with the asymptotic approximation in blue and dashed.}
\label{fig:WF}
\end{figure}

The strategy we have pursued 
is to numerically integrate \eqref{equ:fpWF} starting with initial conditions calculated from
the asymptotic series \eqref{equ:VWFasy} at some large enough $\phi=\phi_\infty>0$ such that
the series is well convergent.\footnote{As we can see from \eqref{equ:VWFasy}
the value of $\phi_\infty$ depends on $A$ but $\phi_\infty=30$ results
in accurate enough initial conditions for values of $A$ as small as $A=10^{-5}$.}
We then record the values of $V_*(0)$ and $V'_*(0)$ as a function of the asymptotic parameter
$A$. The result is shown in fig. \ref{fig:WF}, where we have varied $A$ from $A\to0$ at the 
top end of the curve to $A=50$ at the bottom end of the curve.

In principle (and later we will see in practice) there can exist fixed point potentials with no symmetry under $\phi\leftrightarrow-\phi$. Such solutions would have different values of $A$ in \eqref{equ:VWFasy}, depending on whether $\phi$ is positive or negative. However, since the fixed point differential equation \eqref{equ:fpWF} is symmetric under $\phi\leftrightarrow-\phi$, there would then automatically be two solutions: $V_*(\phi)$ and $V_*(-\phi)$. Therefore solutions with no symmetry can be recognised by the fact that in a plot of $V'_*(0)$ {\it vs.} $V_*(0)$, there are pairs of points with the same $V_*(0)$ but equal and opposite $V'_*(0)$. Fixed point solutions that are odd under $\phi\leftrightarrow-\phi$ would additionally have $V_*(0)=0$ of course, whilst even potentials are associated to only one point since $V'_*(0)=0$.

We can see from the plot in  the left hand panel of 
fig.~\ref{fig:WF}, that there are in fact just two fixed point potentials and both of these are even. The first solution appears at $A=0$ and this corresponds to the Gaussian fixed point
whose exact solution is simply $V_\mathrm{G}(\phi)=1/3$.

The second solution is the Wilson-Fisher fixed point with (to 5 significant figures):
\begin{equation}
\label{equ:WFpars}
 A_{\mathrm{WF}}=0.00100  \qquad \text{and} \qquad V_\mathrm{WF}(0)=0.40953
\end{equation}
and according to the plot in fig. \ref{fig:WF} there are no other  fixed point solutions for larger $A$.
The Wilson-Fisher fixed point potential is plotted on the right in fig. \ref{fig:WF} (black and solid line)
together with the asymptotic approximation given by \eqref{equ:VWFasy} with $A=A_\mathrm{WF}$ (blue and dashed).

We now turn our attention to perturbations around the Gaussian and the Wilson-Fisher fixed point.
Writing
\begin{equation}
 \label{equ:perts}
 V(\phi,t)=V_*(\phi) + \eps v(\phi) \exp(-\lambda t),
\end{equation}
where $V_*$ stands for either of the fixed point potentials and $\eps$ is infinitesimal and dimensionless, and 
remembering that $t=\ln(k/k_0)$
we see that the eigenoperator $v(\phi)$ has the infinitesimal coupling $g=\eps k_0^\lambda$ close to the fixed point
with mass dimension $\lambda$.	
Substituting the ansatz \eqref{equ:perts} into the full flow equation \eqref{equ:flowLPA} with $h=0$ and linearising in $\eps$ we arrive at the eigenoperator equation
\begin{equation}
 \label{equ:WFeops}
 (\lambda-3) v  + \frac{1}{2} \phi \,v' ={\frac {v'' }{ \left( 1+V_*''  \right) ^{2}}}.
\end{equation}
At the Gaussian fixed point, where $V_*''=0$, this eigenoperator equation can be solved exactly and
one recovers the marginal operator $v(\phi)=\phi^6 + subleading$ and  the expected six
relevant operators with eigenvalues $\lambda=1/2,1,\dots,5/2,3$. In addition to this
we find the infinite sequence of irrelevant operators with leading pieces 
 $\phi^7,\phi^8,\dots$ 
corresponding to negative half integer values of $\lambda$.

For the Wilson-Fisher fixed point, $V_*'' \neq0$ and the eigenoperator
equation cannot be solved analytically anymore. In order to find its solutions numerically,
we follow a similar strategy as for the fixed point equation. The first step
consists in developing the large field behaviour of the eigenoperator $v(\phi)$.
In doing so we expand $v(\phi)=v_0(\phi)+v_1(\phi)+\dots$, where $v_0(\phi)$ solves the
left hand side of \eqref{equ:WFeops} and $v_n(\phi)$ is obtained by substituting
$v(\phi)=v_0(\phi)+v_1(\phi)+\dots+v_n(\phi)$  into the left hand side 
and the same expansion of one lower order into the right hand side of \eqref{equ:WFeops}, keeping 
only the leading term of a Taylor expansion in $1/\phi$ of the right hand side and solving the resulting differential equation.
Of course, in this process we have
to use the asymptotic expansion \eqref{equ:VWFasy} of $V_*(\phi)$ to corresponding order.
The first few terms in the resulting asymptotic series for the eigenoperators take the form
\begin{equation}
\label{equ:WFeopsasy}
 v(\phi) = {|\phi|}^{-2\,\lambda+6}
	      -{\frac {1}{4500}}\,{\frac { \left( 2\,\lambda-5\right) \left( 2\,\lambda-6 \right) }{A_\mathrm{WF}^{2}}} {|\phi|}^{-2\,\lambda-4}
	      +{\frac {1}{94500}}\,{\frac { \left( 2\,\lambda-5 \right) \left( 2\,\lambda-6 \right)}{A_\mathrm{WF}^{3}}}{|\phi|}^{-2\,\lambda-8} 
	      +\dots,
\end{equation}
where we have normalised the eigenoperator such that the first term in this expansion has unit coefficient
and $A_\mathrm{WF}$ is given in \eqref{equ:WFpars}.

Using the asymptotic series \eqref{equ:WFeopsasy} for any given $\lambda$ to calculate the initial conditions
at some sufficiently large $\phi_\infty$, we numerically integrate \eqref{equ:WFeops} to determine $v'(0)$. Since the
fixed point potential $V_\mathrm{WF}$ is an even function of $\phi$, the eigenperturbations can be classified as even or
odd functions of $\phi$. To begin with we are interested in the even eigenperturbations, so we impose the condition $v'(0)=0$. 
This will quantise the eigenspectrum
leaving us with a discrete set of eigenoperators. Apart from the vacuum energy $v=1$ with $\lambda=3$
which is clearly an exact solution of \eqref{equ:WFeops}, the only relevant eigenperturbation we find
has the eigenvalue
\begin{equation*}
 \lambda_\mathrm{WF}=1.539,
\end{equation*}
corresponding to a critical exponent $\nu=1/\lambda=0.649$
which is within 3\% of the results obtained at 
$O(\partial^2)$ of the derivative expansion and other approximation methods, such as Monte Carlo methods, resummed perturbative
calculations or the scheme proposed by Blaizot, M\'endez-Galain and Wschebor \cite{Benitez:2011xx}.

For odd eigenperturbations the spectrum will instead be quantised by the condition $v(0)=0$. 
The only two relevant odd eigenoperators are then $v(\phi)=\phi$ with $\lambda=5/2$ which solves the left hand side
of \eqref{equ:WFeops} and $v(\phi)=V_\mathrm{WF}'(\phi)$ with $\lambda=1/2$. The latter however is a redundant
eigenoperator that corresponds to the change of field variable $\phi \mapsto \phi + const.$, \cf sec. \ref{sec:missingvacua}.

\subsection{Cutoff operators with explicit field dependence} \label{sec:gencutoff}
Having confirmed in the previous section that our methods reproduce known results, we are now going
to investigate how the fixed point structure and the critical exponents change if we consider
the flow equation \eqref{equ:flowLPA} with \eqref{cutoff1} and $\al \neq 0$. Apart from a proliferation of fixed points,
we will also encounter a partial breakdown of the LPA in the sense of ref. \cite{DietzMorris:2013-2}.

The fixed point equation we are now interested in solving is obtained from \eqref{equ:flowLPA} by substituting
the dimensionless version of \eqref{cutoff1}, 
\begin{equation} \label{equ:FPgen}
 3V_*- \frac{1}{2} \phi V'_*  = 
 \frac{\left(1-\al\phi^2\right)^{\frac{3}{2}} \left(1- \frac{1}{2}\al \phi^2 \right)}{1-\al \phi^2+V''_*}\,\theta\!\left(1-\al\phi^2\right)
\end{equation}
and we will let $\al$ take positive and negative values in the following.
Due to the fractional exponent on the right hand side however, the analysis proceeds along the same lines
as in the previous section only for $\al<0$. In particular, the asymptotic expansion for large
$|\phi|$ depends on only one free parameter $A$,
\begin{multline}
 V_*(\phi) = A \phi ^{6}+{\frac {1}{150}}\,{\frac {{|\alpha|}^{5/2} \left| \phi \right| }{A}}
 +{\frac {1}{6300}}\,{\frac {{|\alpha|}^{3/2} \left( 105\,A-{\alpha}^{2} \right) }{{A}^{2} \left| \phi \right| }} \\
 +{\frac {1}{486000}}\,{\frac {\sqrt {|\alpha|} \left( 6075\,{
A}^{2}-270\,{\alpha}^{2}A+2\,{\alpha}^{4} \right) }{{A}^{3} 
 \left| \phi \right| ^{3}}} + \dots .
\end{multline}

If $\al>0$ the right hand side of the fixed point equation \eqref{equ:FPgen} becomes zero at $|\phi|=\phi_c=1/\sqrt{\al}$
and vanishes identically for all $|\phi|>\phi_c$. This is a direct consequence of the 
field dependent version
of the optimised cutoff we are using.
For $|\phi|>\phi_c$ the solution of \eqref{equ:FPgen} is therefore simply given by $V_*(\phi)=A\phi^6$
and this solution has to be matched onto the solution obtained from the full fixed point equation for $|\phi|<\phi_c$.
Due to the fractional power on the right hand side of \eqref{equ:FPgen} it is problematic to start numerical integration
directly at the matching point $\phi=\phi_c$. We therefore develop the potential into a generalised Taylor expansion
around the matching point according to
\begin{equation} \label{equ:Vtaylor}
 V_*(\phi)= a_0 + a_1\left(\frac{1}{\sqrt{\al}}-\phi \right)^{\frac{1}{2}} +\frac{a_2}{2!}\left(\frac{1}{\sqrt{\al}}-\phi \right)
 + \frac{a_3}{3!}\left(\frac{1}{\sqrt{\al}}-\phi \right)^\frac{3}{2}+ \frac{a_4}{4!}\left(\frac{1}{\sqrt{\al}}-\phi \right)^2 + \dots.
\end{equation}
To do this in practice it is convenient to perform the change of variable $\phi \mapsto 1/\sqrt{\al}-u^2$
before substituting this expansion into \eqref{equ:FPgen}\footnote{The form of the generalised
Taylor expansion \eqref{equ:Vtaylor} is dictated by the fractional power in \eqref{equ:FPgen}.}.  For the fixed point equation to be 
satisfied order by order in $u$ we then find that $a_0$ and $a_2$ are free parameters, corresponding
to two initial conditions at the matching point, whereas $a_1=a_3=0$ and all higher coefficients 
$a_4,a_5,\dots$ are functions of $a_0$ and $a_2$. Since we want to match the solution 
\eqref{equ:Vtaylor} at $\phi=1/\sqrt{\al}$ to the asymptotic solution $V_*(\phi)=A\phi^6$ we choose the initial
conditions accordingly:
\begin{equation}\label{equ:a0a2}
a_0=\frac{A}{\al^3} \qquad \text{and} \qquad a_2 = -\frac{12A}{\al^\frac{5}{2}}.
\end{equation}
In this way,\footnote{Actually, it is convenient to make 
the identification \eqref{equ:a0a2}  before substituting \eqref{equ:Vtaylor} into
\eqref{equ:FPgen} to avoid the appearance of vanishing denominators at intermediate stages.} the higher coefficients $a_4,a_5,\dots$ become functions of $A$ and $\al$,
\begin{equation} \label{asyexpcoeffs}
a_{{4}}=360\,{\frac {A}{{\alpha}^{2}}}, \qquad a_{{5}}={\frac {16\sqrt {2}}{5}}\,\frac{{
\alpha}^{{\frac {13}{4}}}}{{A}}, \qquad a_{{6}}=-{\frac {8}{75}}\,{
\frac {135000\,{A}^{4}+{\alpha}^{10}}{{\alpha}^{3/2}{A}^{3}}}, \dots
\end{equation}
We note that this form of $a_4$ leads to automatic matching of the second derivative $V''_*(\phi)$
at $\phi=\phi_c$ but higher derivatives of the potential will diverge as we approach
the matching point from the left. This is clearly an artefact of the cutoff choice 
and does not have any physical significance. We will see now that despite this non-smoothness of the potential
the overall picture emerging from our analysis is consistent.

\begin{figure}[ht]
  \begin{center}
  $
   \begin{array}{ccccc}
     \includegraphics[width=0.3\textwidth,height=0.225\textheight]{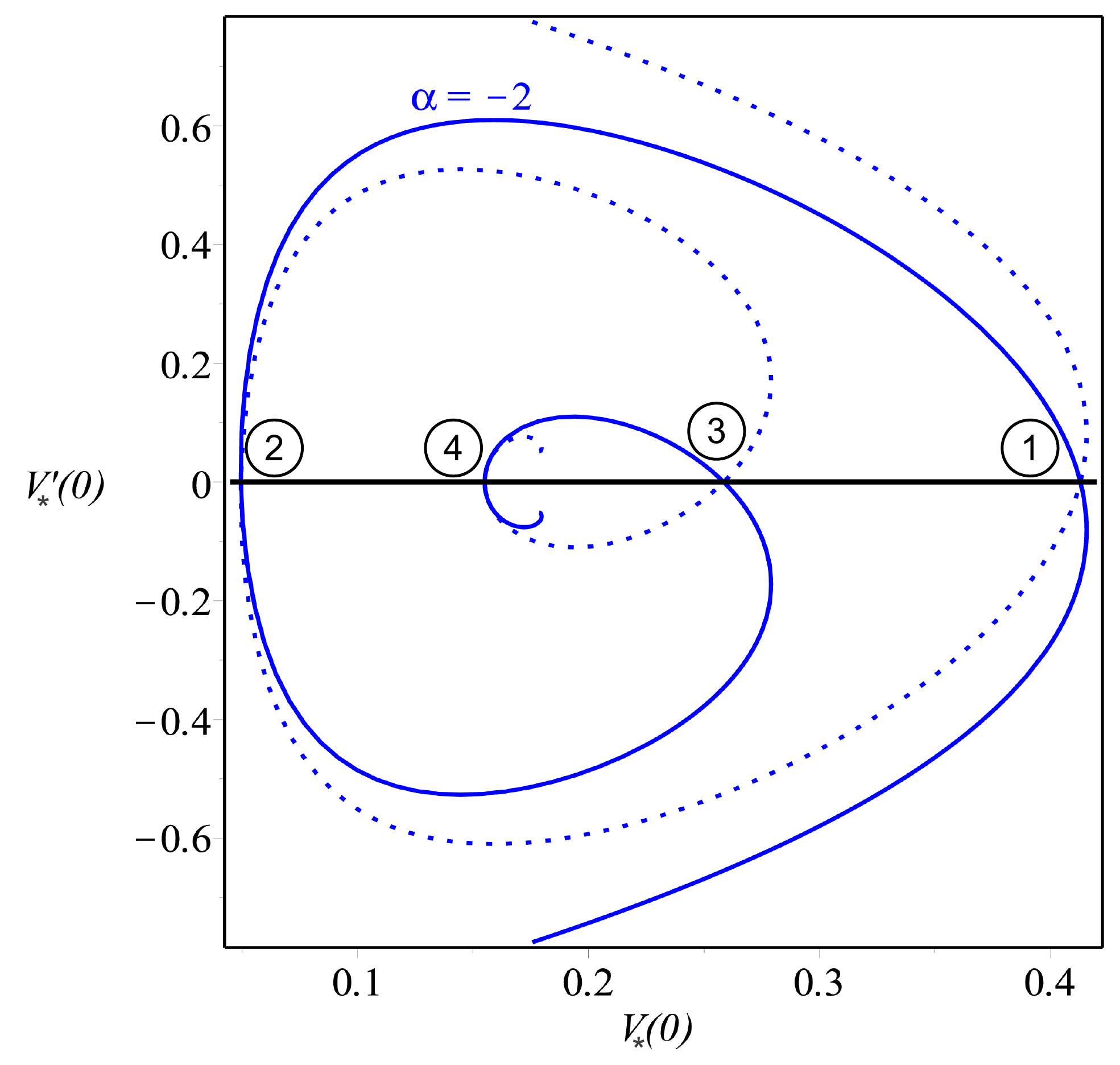}  & &
     \includegraphics[width=0.3\textwidth,height=0.225\textheight]{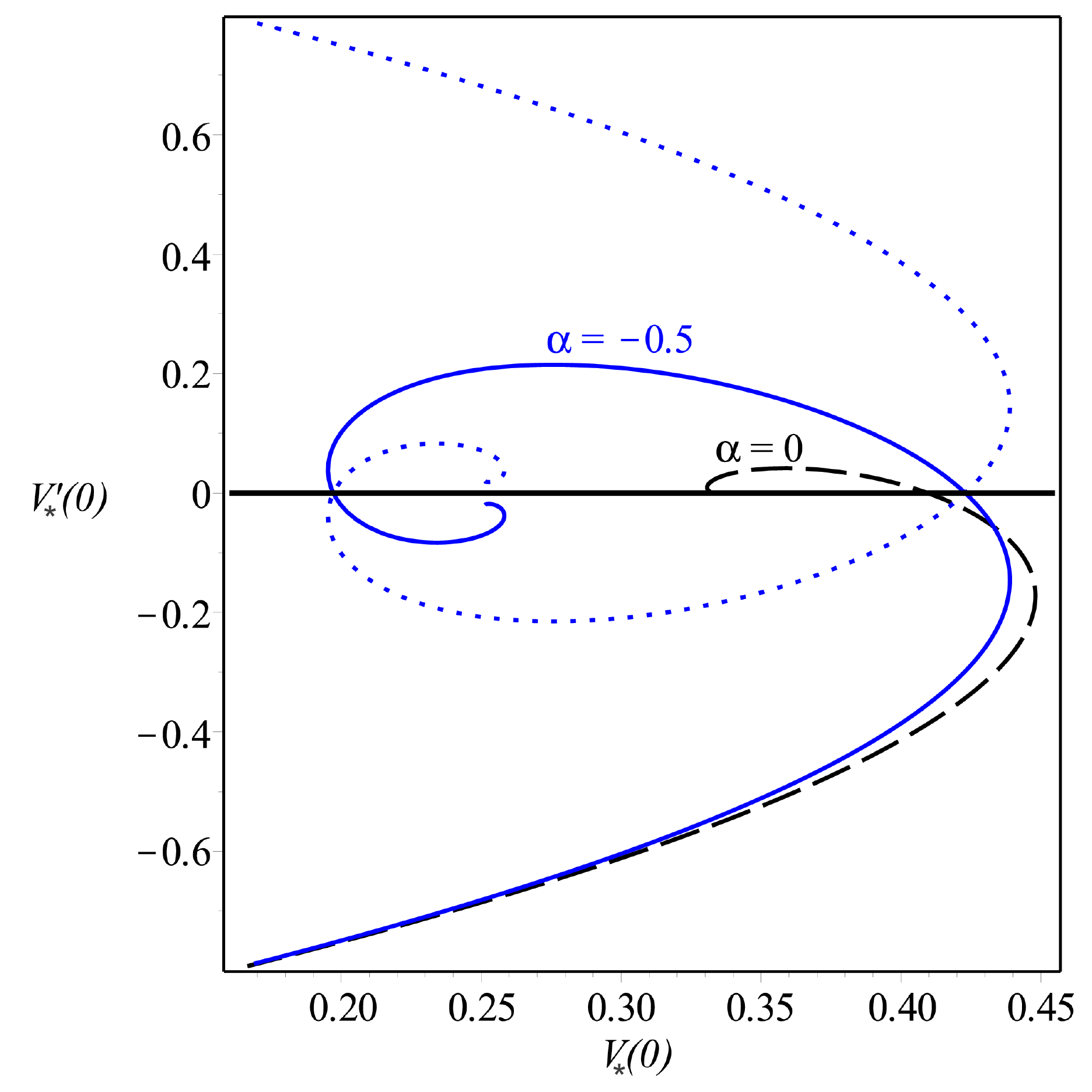} && 
     \includegraphics[width=0.3\textwidth,height=0.225\textheight]{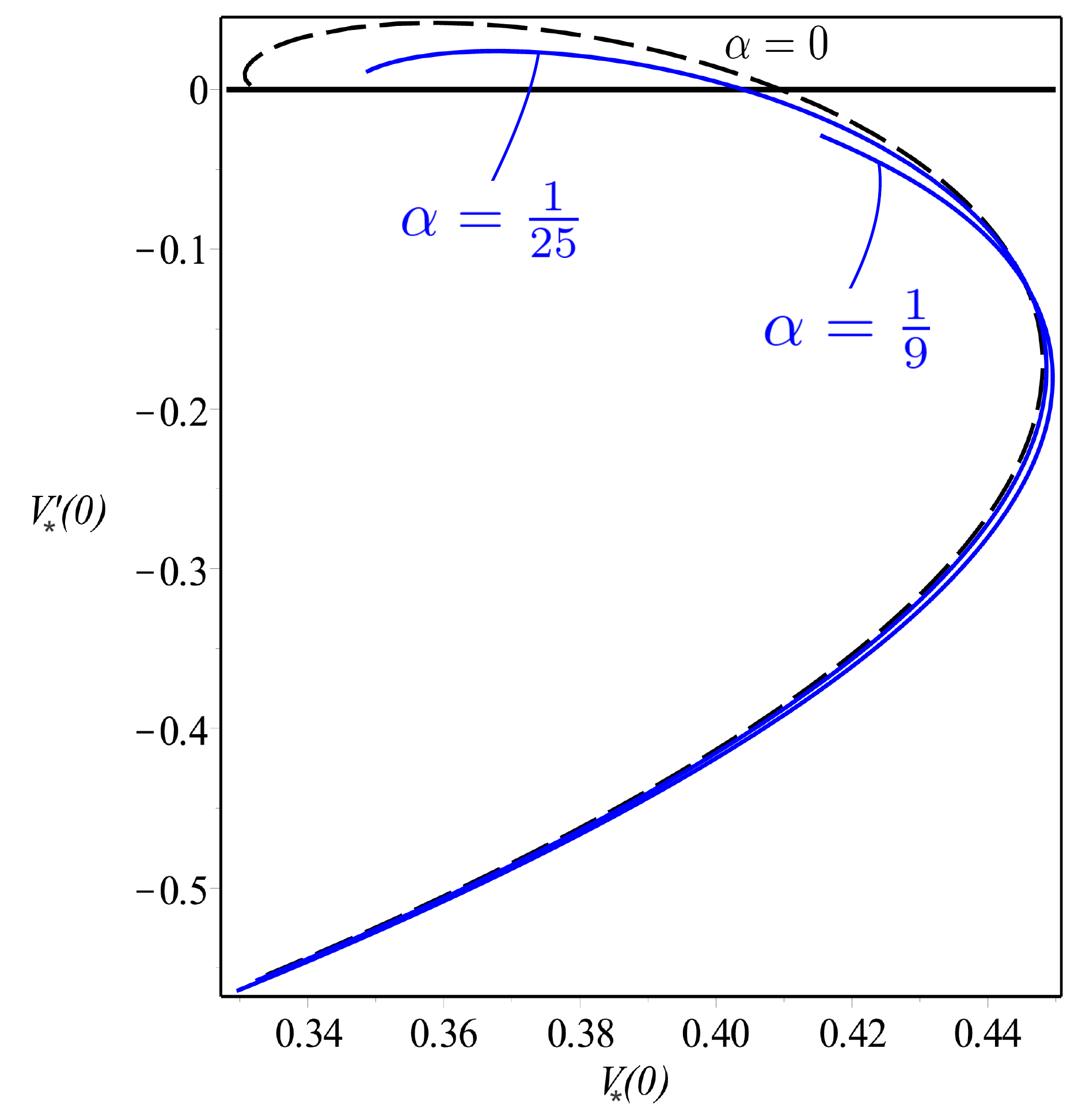}
   \end{array}$
  \end{center}
   \caption{Varying the asymptotic parameter $A$ for different values of $\al$ leads to these spirals in $V_*(0),V'_*(0)$ - space.
   In dashed and black we have reproduced the case $\al=0$ from fig. \ref{fig:WF} whereas the blue, solid curves belong to various
   values of $\al \neq0$. The blue dotted curves correspond to the blue curves reflected in the $V_*(0)$ axis. From the discussion below \eqref{equ:VWFasy}, we know that fixed point solutions correspond to the points where the blue curves and dotted-blue curves cross. }
\label{fig:gen}
\end{figure}

After these preparations and just like in the case of $h=0$ in the previous section
 we now vary the asymptotic parameter $A$ and numerically integrate
the fixed point equation \eqref{equ:FPgen} to find $V_*(0)$ and $V'_*(0)$ as functions of $A$. The result is shown in
fig. \ref{fig:gen} for various positive and negative values of $\al$. 

Let us first discuss the two plots pertaining to $\al<0$. As soon as $\al$ becomes negative,
what was the Gaussian fixed point at the top end of the curve corresponding to $A=0$, drops below the $V_*(0)$ - axis
and thus it is no longer present. Instead a new fixed point with non-constant potential comes into being which for $\al=-0.5$
is located at $V_*(0)\approx0.2$. Close to where the Wilson-Fisher fixed point was for $\al=0$ we still find a second
fixed point and we will see that it still has only one relevant even eigenperturbation
(ignoring the vacuum energy)  like the original Wilson-Fisher fixed point.

As we decrease $\al$ further, more and more fixed points are created. On the left in fig. \ref{fig:gen} we see easily that 
for $\al=-2$ there are already four fixed points,
the one on the far right corresponding to the Wilson-Fisher fixed point. These fixed points are labelled \tc1 to \tc4 in the direction of decreasing $A$.

Less easy to see, even given the dotted blue curve to guide the eye, is the pair of points $V_*(0),\pm V'_*(0)$ close to \tc4, 
corresponding to the first non-symmetric fixed point pair. As $\al$ is decreased still further these non-symmetric solutions 
become more visible and are joined by both more symmetric fixed point solutions and other pairs of non-symmetric fixed point solutions.

On the other hand, if we go to $\al>0$ the spiral from $\al=0$ progressively unwinds. We immediately lose the Gaussian
fixed point but retain the Wilson-Fisher fixed point. Increasing $\al$ further however leaves us with no fixed points 
at all.

In order to understand the nature of the new fixed points and to justify that, provided $\al$ is not too large and positive,
an image of the Wilson-Fisher fixed point is still present, we have to investigate their eigenspectra.
The eigenoperator equation takes the form
\begin{equation}
 \label{equ:eopsgen}
  (\lambda-3) v  + \frac{1}{2} \phi \,v' =  
  \frac{\left(1-\al\phi^2\right)^{\frac{3}{2}} \left(1- \frac{1}{2}\al \phi^2 \right)}{\left(1-\al \phi^2+V_*''\right)^2}v''(\phi)
  \, \theta \! \left(1-\al \phi^2 \right).
\end{equation}
For each fixed point potential $V_*$ this equation can be analysed with the same methods as in sec. \ref{sec:WF} provided
$\al<0$. For $\al>0$ the right hand side of \eqref{equ:eopsgen} again 
 vanishes for $|\phi| \geq \phi_c=1/\sqrt{\al}$ and
we have to match the solution in the range $\phi<\phi_c$ to the solution $v(\phi) = \phi^{6-2\lambda}$ of the left hand
side at $\phi=\phi_c$. This is done along the same lines as for the fixed point equation \eqref{equ:FPgen}
via a generalised Taylor expansion of the eigenoperator $v$ at the matching point.

In the following we will focus our attention on the even fixed points in fig. \ref{fig:gen}, leaving aside the two non-symmetric
solutions close to fixed point \tc4. This in turn allows us to search for even eigenoperator solutions of \eqref{equ:eopsgen} with
the aim of comparing the results to the single even eigenperturbation found for the Wilson-Fisher fixed point in the previous section.

As labelled in fig.~\ref{fig:gen}, we then find that the $n$-th fixed point has $n$ relevant even eigenoperators
if we do not take into account the vacuum energy.
In particular we see that the first fixed point always has one relevant
even eigendirection
which justifies regarding it 
as the Wilson-Fisher fixed point. Of course, the eigenvalue associated to this relevant direction
depends on $\al$ as can be seen from table \ref{tab:evWF}.
It decreases along with $\al$ and we have checked that it is still relevant at $\al=-8$ which
suggests that it may	 slowly tend to zero from the right as $\al \rightarrow -\infty$.
In table \ref{tab:evs} we give the eigenvalues 
pertaining to the relevant even eigenoperators
of the additional new fixed points
labelled in fig. \ref{fig:gen} on the left.
\begin{table}[h]
\parbox{0.45\textwidth}{
\centering
  \begin{tabular}{|c|c|c|c|c|}
   \hline $\al$ & $1/25$ & 0 & $-0.5$ & $-2$ \\
   \hline $\lambda_\mathrm{WF}$ & 1.62 & 1.54 & 1.17 & 0.89 \\
   \hline
  \end{tabular}
\caption{
The eigenvalue of the single relevant even eigendirection of the Wilson-Fisher fixed point 
for the values of $\al$ used in fig. \ref{fig:gen}.}
\label{tab:evWF}
  }
  \hspace{1cm}
 \parbox{0.45\textwidth}{
 \centering
\begin{tabular}{|c|c|c|c|c|}
  \hline
   FP & $\lambda_1$ & $\lambda_2$ & $\lambda_3$ & $\lambda_4$ \\
  \hline 2 & 2.35 & 0.76 & - & - \\
  \hline 3 & 2.02 & 1.43 & 0.60 & - \\
  \hline 4 & 2.10 & 1.69 & 1.08 & 0.39 \\
  \hline
  \end{tabular}
\caption{Each new fixed point possesses one more relevant even eigendirection. These are the eigenvalues for the labelled
additional fixed points in the plot on the left in fig. \ref{fig:gen}.}
\label{tab:evs}
}
\end{table}

Overall, we conclude that as soon as $\al \neq 0$ we encounter significant differences to the standard
picture of the fixed point structure for a three-dimensional single scalar field. For appropriate values of $\al$ spurious fixed points can be created,
previous fixed points can be lost and the number of possible relevant
even eigendirections changes with $\al$.
More than that, these are not the only problems as we will see in the next section.

\subsubsection{Missing vacua and redundant operators}
\label{sec:missingvacua}
Redundant operators are eigenoperators which are equivalent to an infinitesimal change
of field variable and do not have a well defined renormalisation group eigenvalue \cite{WR,DietzMorris:2013-2}.
Thus they are not part of the true eigenspace of physical eigenperturbations.
It was shown recently that for one of the flow equations
in the $f(R)$ approximation in quantum gravity, the eigenspaces of all fixed points collapse to a point
as a consequence of the fact that all eigenoperators are redundant, which in turn is a result of the equations of motion possessing no vacuum solution in their domain of definition \cite{DietzMorris:2013-2}. In the following we show that a 
version of this pathology also occurs in the present context when $\al \neq 0$.
To this end we remark that even though we have so far been concerned with even eigenoperators only,
we now include odd eigenoperators in the discussion below.

At the level of the LPA in scalar field theory, an eigenoperator $v$ is redundant
if it can be written in the form
\begin{equation} \label{equ:redcond}
v(\phi)=\zeta(\phi) V_*'(\phi),
\end{equation}
where we are free to choose the function $\zeta$ with the only requirement that it has to
be non-singular on its domain of definition $\phi \in (-\infty , \infty)$ \cite{DietzMorris:2013-2}.
This implies that turning points of the fixed point potential $V_*$ are zeros of the eigenoperator.

\begin{figure}[h]
\centering
 \includegraphics[width=0.5\textwidth,height=0.35\textheight]{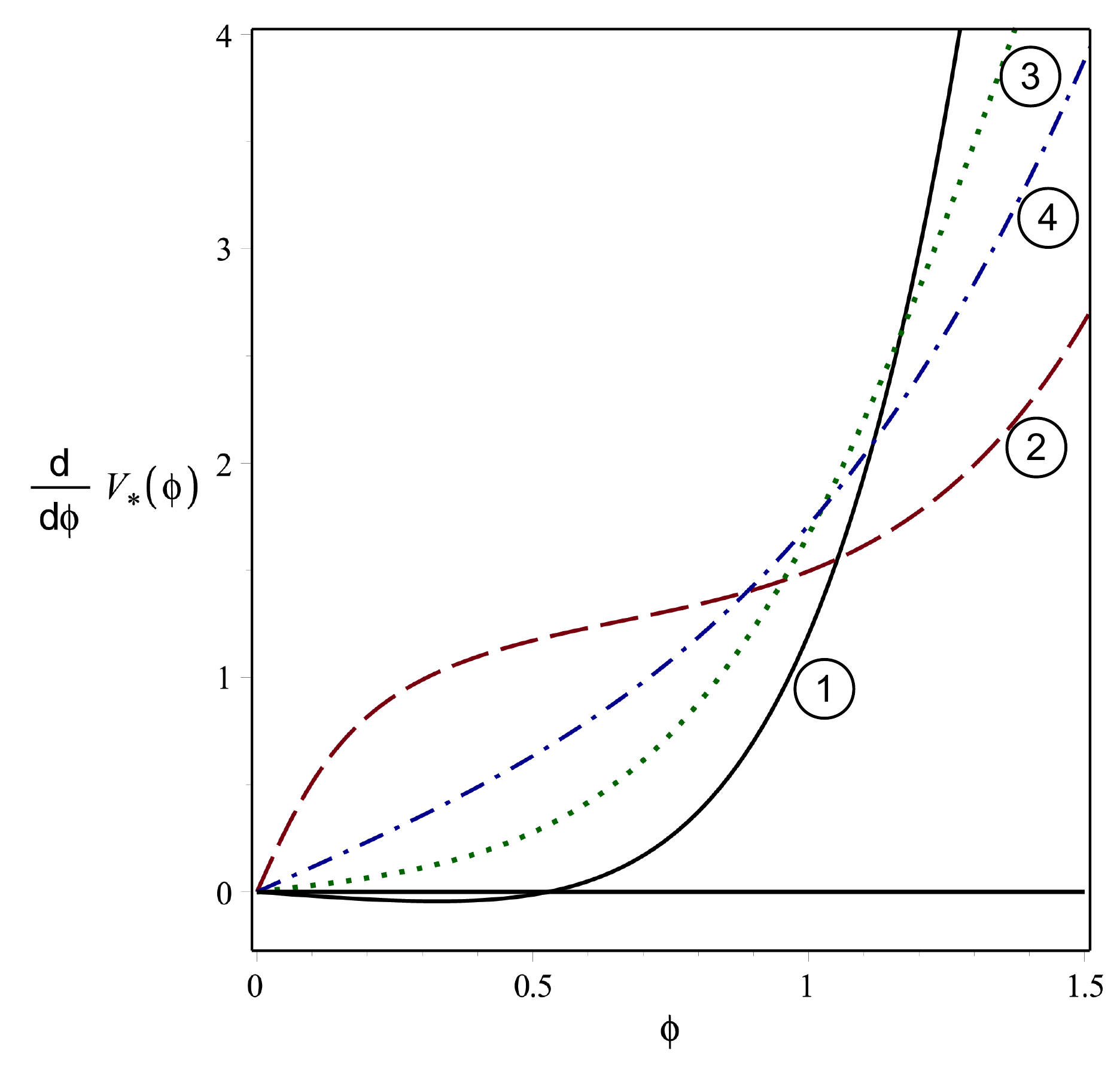}
 \caption{The derivative of all four fixed point potentials from fig. \ref{fig:gen} (left).
 Only the Wilson-Fisher fixed point features a turning point at non-zero field.}
 \label{fig:vps}
\end{figure}

Considering the Wilson-Fisher fixed point (with arbitrary $\al$) we always find 
a non-trivial turning point $V_\mathrm{WF}'(\pm \phi_0)=0$ at some $\phi_0 > 0$. For $\al=0$ this can be
seen from fig. \ref{fig:WF} (right) and for $\al=-2$ we have plotted $V_\mathrm{WF}'$ in fig. \ref{fig:vps}.
Given that the eigenoperator
equation \eqref{equ:eopsgen} is already constrained such that it gives rise to a quantised
eigenspectrum, the additional condition $v(\phi_0)=0$ implied by \eqref{equ:redcond}
overconstrains the system and we will find no solutions at all.
In the case of the Wilson-Fisher fixed point we have explicitly checked for the values of $\al\neq0$ given in fig. \ref{fig:gen}
that $v(\phi_0)\neq 0$ is indeed the case for all
relevant eigenoperators, both odd and even. Hence \eqref{equ:redcond} cannot hold
for a non-singular function $\zeta$ and none of its eigenoperators are redundant.
For $\al=0$ there is one exception, given
by the redundant odd eigenoperator $v=V_\mathrm{WF}'$ with $\lambda=1/2$. This is due to
the shift symmetry $\phi(x) \mapsto \phi(x) +\delta$ of the unscaled equation \eqref{equ:FRGE} with a field independent cutoff,
\footnote{{\it N.B.} This is a symmetry of the \emph{flow equation} and this should not be confused with any symmetries of the fixed point solutions. Moreover this symmetry becomes $t$-dependent once we adopt dimensionless variables and hence is not a symmetry of the 
scaled equation \eqref{equ:fpWF}.}
\cf \cite{DietzMorris:2013-2,Morris:1994jc}.

The state of affairs is different for all the other labelled fixed points in fig. \ref{fig:gen} (left).
The only turning point for these fixed points is at $\phi=0$ which is a consequence of their $Z_2$-symmetry, \cf fig. \ref{fig:vps}.
Even eigenperturbations, \ie eigenperturbations satisfying $v'(0)=0$, are still not redundant as they must have $v(0) \neq 0$ 
if they are not to vanish identically. 
However, since odd eigenoperators vanish at $\phi=0$ they satisfy $v'(0)\neq0$ and hence would 
be redundant if the fixed point potential does not vanish faster than linearly, \ie if $V_*''(0)\neq 0$.
From the fixed point equation \eqref{equ:FPgen} we see that $V_*''(0)= 0$ implies $V_*(0)=1/3$.
Together with $V'_*(0)=0$ these unique initial conditions will only define a fixed point solution 
for a discrete set of values for $\al$ (\eg if $\al=0$ they describe the Gaussian fixed point)
but can be excluded otherwise. We therefore see that generically all odd eigenperturbations
of all fixed points other than the Wilson-Fisher fixed point are redundant if $\al<0$.

As mentioned in the Introduction, we will resolve all problems that have appeared here and in the previous section
and recover
the correct description of three-dimensional single component scalar field theory with the cutoff choice
\eqref{cutoff1} in \eqref{equ:cutoffh} by exploiting the modified shift Ward identity in sec. \ref{sec:sWI-LPA}.

\subsection{A cutoff operator depending on the potential}\label{sec:LPALitim}
The results of the previous section were obtained with the purely explicit field dependent choice
\eqref{cutoff1} for the function
$h$ in the cutoff operator \eqref{equ:cutoffh}. As a second example, we now consider the choice \eqref{cutoff2} which,
contrary to \eqref{cutoff1},
has occurred previously in the literature \cite{Litim:2002hj}. It carries intrinsic scale dependence from the field dependent propagator in \eqref{equ:FRGE},
\ie the second functional differential of equation \eqref{equ:LPA}. 
We have seen at the end of sec. \ref{sec:LPAsetup} that the correct one-loop $\beta$ function in four dimensions is then maintained for any $\alpha$. 
One might hope, with the background field dependence of the cutoff being thus adapted to   \eqref{equ:FRGE} in this way 
that at least qualitatively unphysical behaviour would now be avoided. 
However we will see in this section that that hope is unfounded.

Using \eqref{cutoff2} in the general expression \eqref{equ:flowLPA}, the flow equation in dimensionless
variables becomes
\begin{equation}
\label{equ:V''Flow}
\partial_{t}V-\frac{1}{2}(d-2)\phi V'+dV=\dfrac{\big(1-\al(1+\frac{1}{2}\partial_{t})V''+\frac{1}{4}\al(d-2)
\phi V'''\big)\big(1-\al V''\big)^{d/2}}{\left(1+(1-\al)V''\right)}\theta(1-\al V'').
\end{equation}
We emphasise again that the background field from \eqref{equ:cutoffh} has disappeared from the flow equation \eqref{equ:V''Flow}
due to the single field approximation as described below \eqref{approx}.	
In \cite{Litim:2002hj} the choice $\al =1$ was considered and we see that this leads to a significant
simplification in the flow equation as the denominator on the right hand side becomes equal to one.
The fixed point equation in three dimensions that we wish to solve is now given by:
\begin{equation}
\label{equ:V''Fixed}
3V_{*}-\frac{1}{2}\phi V_{*}'=\dfrac{(1-\al V_{*}''+\frac{1}{4}\al\phi V_{*}''')(1-\al V_{*}'')^{3/2}}{(1+(1-\al)V_{*}'')}\theta(1-\al V_{*}'').
\end{equation}
The analysis of this equation presents many similarities to the treatment of \eqref{equ:FPgen} in the previous
section. At the same time an important subtlety arises from the appearance of the potential as the function
that is being solved for in the step function on the right hand side.

For large field \ensuremath{\phi} we expect the quantum corrections
from the right hand side of the fixed point equation
\eqref{equ:V''Fixed} to be negligible.  Solving the left hand side gives the asymptotic solution
\ensuremath{V_{*}(\phi)=A\phi^{6}}.  Thus if
\ensuremath{\al>0} the right hand side of this fixed point
equation is zero down to a critical field given by
\ensuremath{V_{*}''(\phi_c)=1/\al}, where we can always assume $\phi_c>0$ due to the symmetry
$\phi \mapsto -\phi$ of the fixed point equation.
This solution has to be matched to the solution obtained from
the full fixed point equation for all
\ensuremath{\lvert\phi\rvert\leq\phi_{c}}, \cf sec. \ref{sec:gencutoff}.  Due to the fractional
power on the right hand side of \eqref{equ:V''Fixed} we proceed
using a Taylor expansion around \ensuremath{\phi_c}:
\begin{equation}
\label{equ:V''Taylor}
V_{*}(\phi)=b_0 + b_1\left(\phi_c-\phi \right)
 +\frac{b_2}{2!}\left(\phi_c-\phi \right)^2
 + \frac{b_3}{3!}\left(\phi_c-\phi \right)^3+
 \frac{b_4}{4!}\left(\phi_c-\phi \right)^4 + \cdots.
\end{equation}
To do this we implement the change of variable
\ensuremath{\phi\mapsto\phi_c-u} in \eqref{equ:V''Fixed}.  For the fixed point equation to
be satisfied order by order in $u$ we then find that
\ensuremath{b_0},  \ensuremath{b_1} and \ensuremath{b_2} are free
parameters, corresponding to the initial conditions set at the
matching point, whilst all higher coefficients depend on these.
Using the argument of the Heaviside function in
\eqref{equ:V''Fixed} and our asymptotic solution we find that we
must match the Taylor expansion at
\ensuremath{\phi=\phi_c=(30A\al)^{-1/4}}.  Choosing the first
three coefficients accordingly:
\begin{equation*}
b_{0}=A\phi_{c}^{6}=A(30A\al)^{-3/2},\qquad b_{1}=-6A\phi_{c}^{5}=-6A(30A\al)^{-5/4},\qquad b_{2}=30A\phi_{c}^{4}=\frac{1}{\al}.
\end{equation*}
This ensures \ensuremath{V_{*}}, \ensuremath{V_{*}'} and \ensuremath{V_{*}''} match smoothly across \ensuremath{\phi_{c}} and we expect all higher coefficients to depend on these three.
Note that we explicitly require matching of the second derivative as we are dealing with a third
order differential equation, in contrast to the matching in the previous section.

Care must be taken as on the right hand side of
\eqref{equ:V''Fixed} the bracket with half integer power contains a
derivative of the potential.  Rewriting the Taylor series as a sum
of initial conditions and a polynomial of higher order terms,
\begin{equation}
\label{equ:TrunTaylor}
V_{*}(u)=A(30A\al)^{-3/2}-6A(30A\al)^{-5/4}u+\frac{1}{2\al}u^{2}+F(u),
\end{equation}
the new Taylor expansion gives \eqref{equ:V''Fixed} in terms of the function of higher orders \ensuremath{F(u)\sim\mathcal{O}(u^{3})}:
\begin{multline}
\label{equ:V''TrunFixed}
\frac{1}{A\al^{2}}\bigg(\big((30A\al)^{3/4}-(30A\al) u\big)F'(u)+180A\al F(u)+60Au^{2}\bigg)\\
=\frac{\al^{9/4}}{2A^{1/4}}\dfrac{\bigg(\big(30(A\al)^{1/4}u-30^{3/4}\big)F'''(u)-120(A\al)^{1/4}F''(u)\bigg)\bigg(-F''(u)\bigg)^{3/2}}{\bigg(1-\al(\al-1)F''(u)\bigg)}.
\end{multline}
We find that due to the bracket with half integer power on right hand side of equation \eqref{equ:V''TrunFixed}
we cannot find a physical fixed point solution if \ensuremath{F(u)} is a 
series with integer powers $n\ge3$. 
To proceed we factor out the leading power of \ensuremath{F(u)} and assume the remaining dependence follows
the form of a generalised power series similar to equation \eqref{equ:Vtaylor} such that:
\begin{equation}
\label{equ:F(u)gammaf}
F(u)=u^{\gamma}\cdot f(u),
\end{equation}  
where \ensuremath{\gamma\geq 3} and the function \ensuremath{f(u)} is a generalised Taylor series.
Expanding equation \eqref{equ:V''TrunFixed} in terms of this ansatz we find that to match both sides order by order in $u$
we require \ensuremath{\gamma=16/5} and the generalised Taylor series \ensuremath{f(u)} must take the form:
\begin{equation}
\label{equ:ftaylorseries}
f(u)=c_{0}+c_{1}u^{1/5}+c_{2}u^{2/5}+c_{3}u^{3/5}+\cdots,
\end{equation}
where we have absorbed combinatoric factors into the coefficients \ensuremath{c_{i}}. 
We may simplify the matching by multiplying through by the denominator on the right hand
side of equation \eqref{equ:V''TrunFixed} such that we only need to series expand the bracket with half
integer power to properly match terms order by order in $u$.
We note that as we are solving a third order equation we should expand the general Taylor
series ansatz in equation \eqref{equ:ftaylorseries} to at least three orders beyond the highest order of $u$ 
we would like to compare coefficients of.
We then find the coefficients for the generalised Taylor series are given by:
\begin{align*}
c_{0}=-\dfrac{25}{88}\left(\dfrac{25}{72}\alpha^{-4}\phi_c^{-2}\right)^{\frac{1}{5}},
\quad c_{1}=\dfrac{125}{5984}\left(\dfrac{625}{162}\alpha^{-13}\phi_{c}\right)^{\frac{1}{5}},\quad 
c_{2}=-\dfrac{71875}{61611264}\left(\dfrac{5}{48} \alpha^{-17}\phi_{c}^{4}\right)^{\frac{1}{5}},
\end{align*}
where for brevity we have only displayed the first few coefficients here.  

We note that the structure of the matching obtained will cause \ensuremath{V_{*}'''(\phi_c)=0} and all higher
derivatives to diverge as we approach the matching point from the left and that, as we noted with equation \eqref{equ:Vtaylor},
this is an artefact of the cutoff chosen and has no physical significance.  
We find that if we use the Taylor expansion to calculate \ensuremath{V_{*}(\phi)\text{, } V_{*}'(\phi) \text{ and } V_{*}''(\phi)} 
at a relative distance of $10^{-7} \phi_{c}$ to the left of \ensuremath{\phi_{c}} 
we may avoid the singularities in the higher derivatives and proceed to numerically integrate the fixed point equation \eqref{equ:V''Fixed}
as we vary the asymptotic parameter A to find $V_{*}$ and $V'_{*}$ at $\phi=0$, \cf sec. \ref{sec:gencutoff}.
In principle, we would also have to record $V''_*(0)$ in order to have a
complete set of initial conditions. As discussed in sec. \ref{sec:WF}, a global solution would then be given by two sets of such initial conditions,
where the components $V_*(0)$ are equal in each set and likewise $V_*''(0)$, whilst $V_*'(0)$ is equal and opposite. As we will see in the following, we can
safely disregard the $V_*''(0)$-component in this procedure as we do not find any two pairs $\left(V_*(0),V_*'(0)\right)$ with
the same $V_*(0)$ and equal and opposite
$V_*'(0) \neq0$. If that was the case, it would be necessary to check that also the second derivative at $\phi=0$ agrees for these pairs.
In other words, since we only find solutions satisfying $V_*'(0)=0$, implying that they are symmetric under $\phi \mapsto -\phi$,
the same value for the asymptotic parameter $A$ can be used to describe the behaviour at $\phi \rightarrow \pm\infty$.

\begin{figure}[ht]
  \begin{center}
  $
   \begin{array}{ccccc}
     \includegraphics[width=0.3\textwidth,height=0.225\textheight]{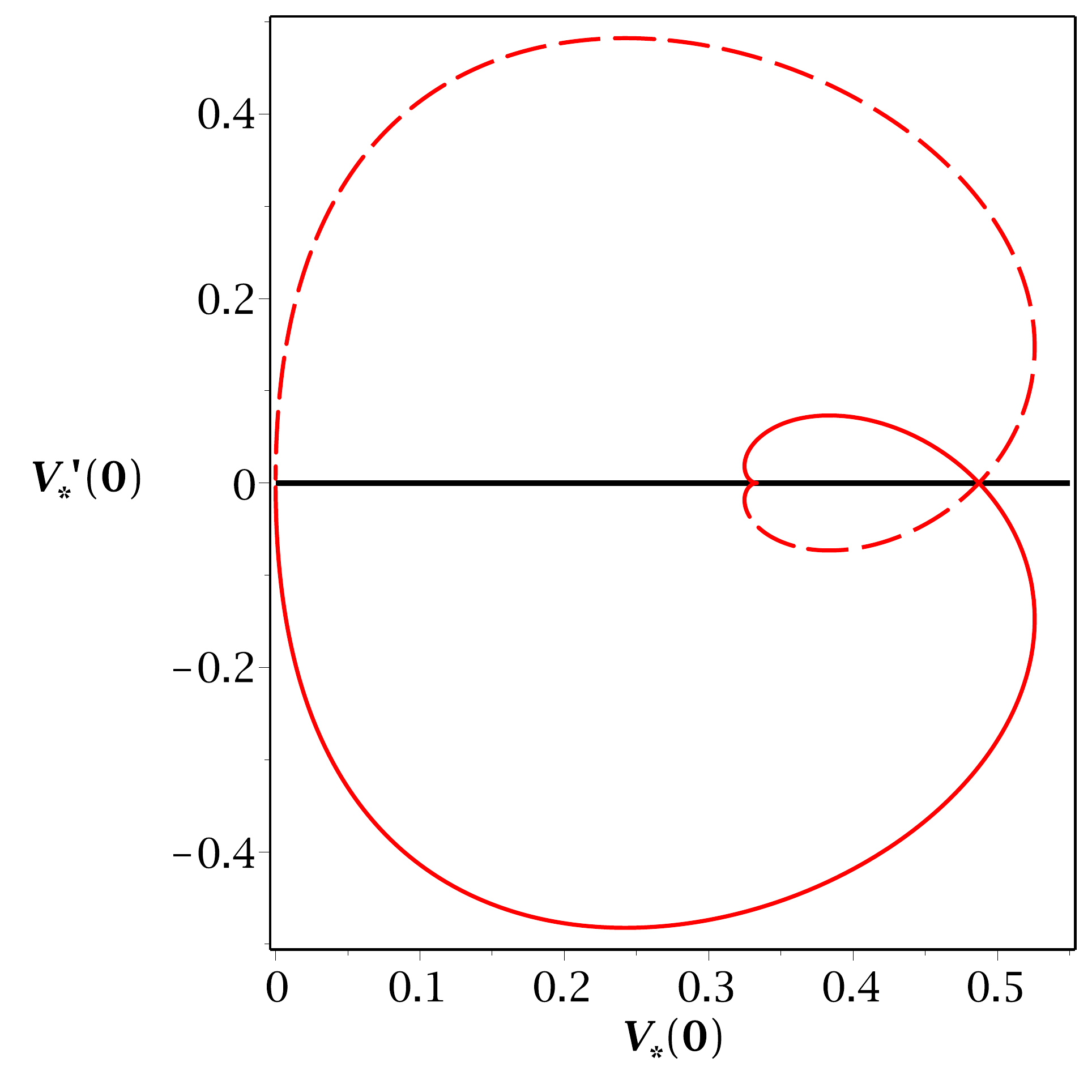}  & &
     \includegraphics[width=0.3\textwidth,height=0.225\textheight]{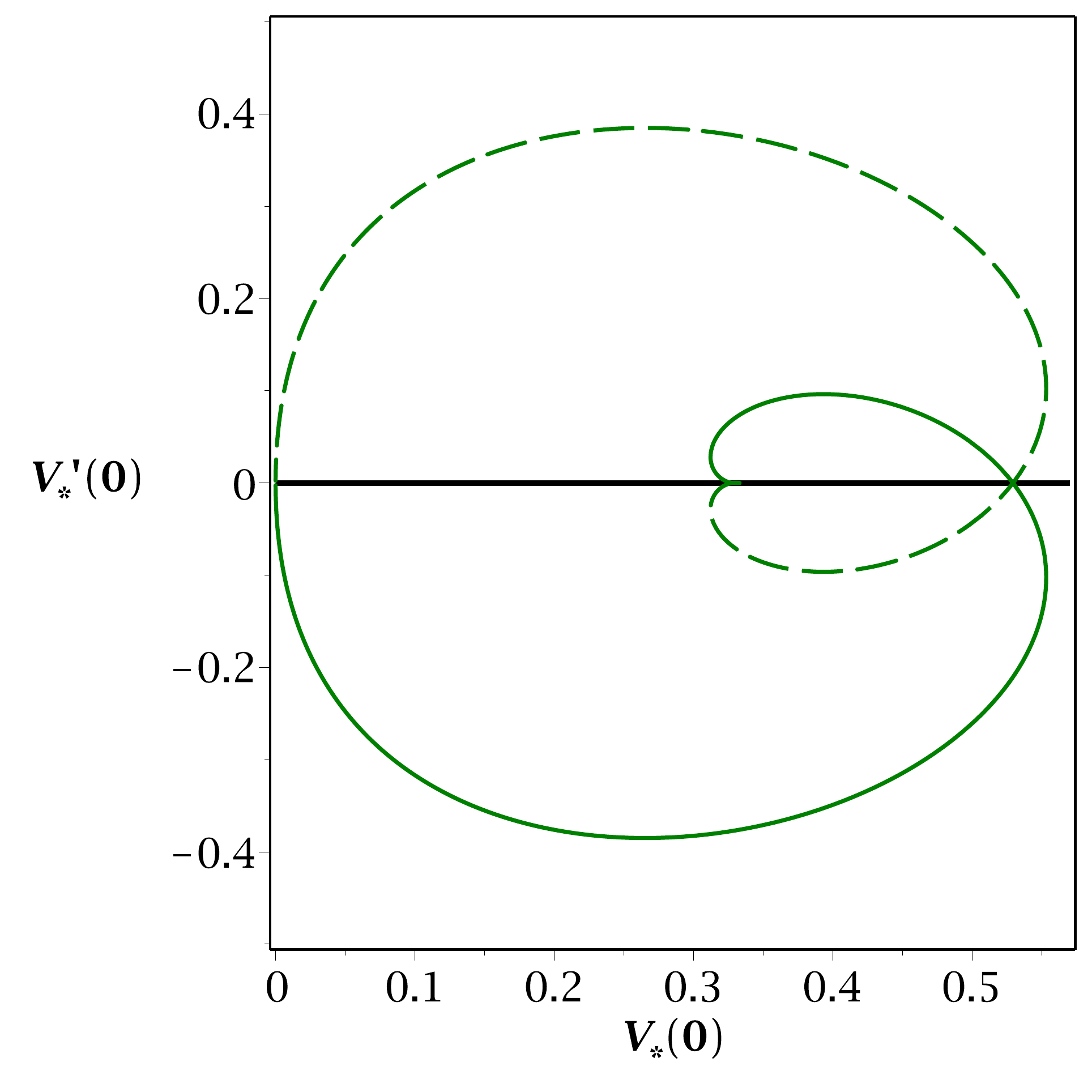} && 
     \includegraphics[width=0.3\textwidth,height=0.225\textheight]{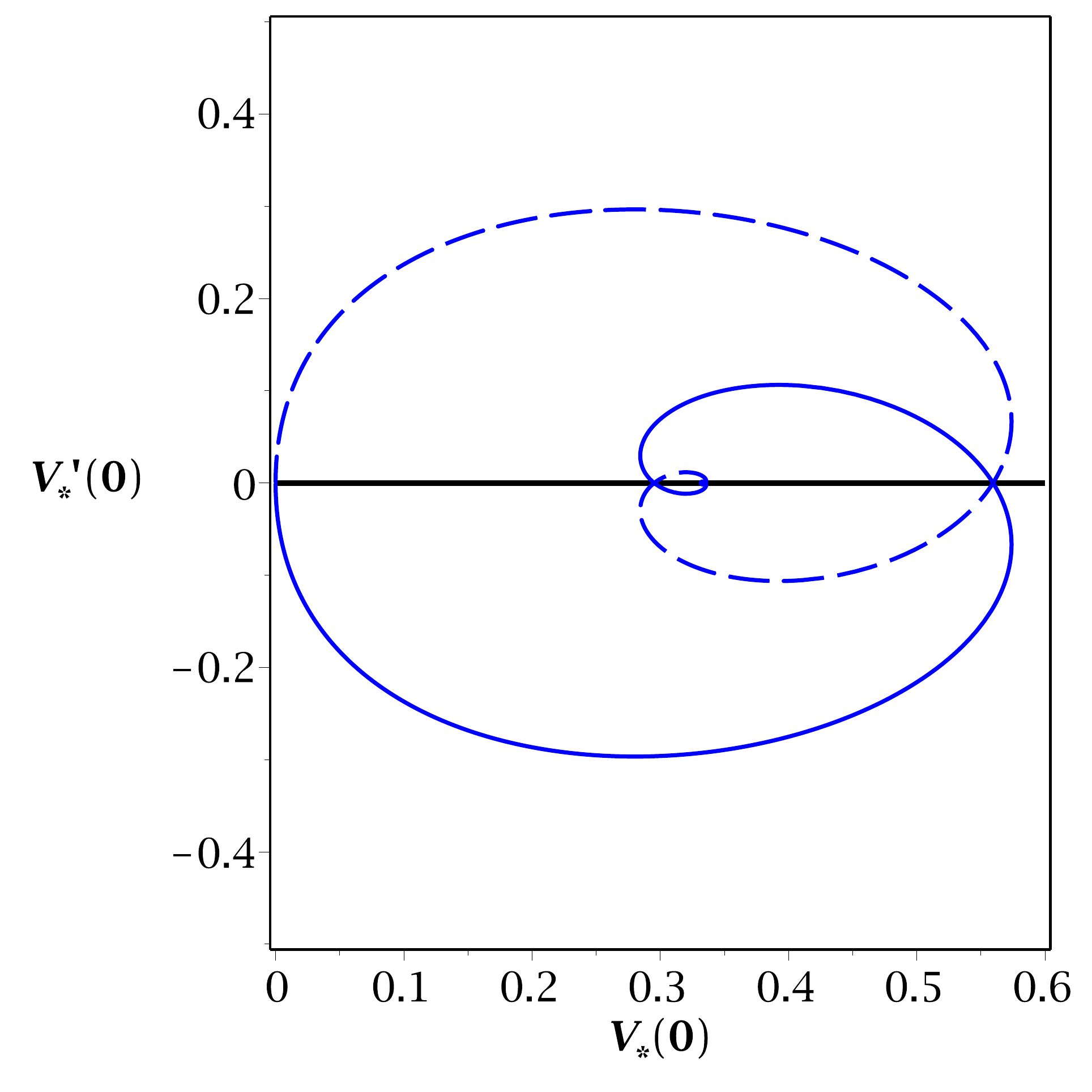}
   \end{array}$
  \end{center}
   \caption{Varying the asymptotic parameter $A$ from $A=0$ to $A\to\infty$, for different values of $\al$ 
   leads to these spirals in $V_*(0),V'_*(0)$ - space, where  $A=0$ corresponds to the point \ensuremath{(V'_{*}(0)=0, V_{*}(0)=1/3)},
   and  $A\to\infty$ tends to the origin.  From left to right the curves show
   \ensuremath{\al=0.5\text{ (red), }\al=1\text{ (green) and } \al=2} (blue). 
   The dashed curves correspond to the solid curves reflected in the $V_*(0)$ axis. 
   From the discussion below \eqref{equ:VWFasy}, we know that fixed point solutions
   correspond to the points where the solid and dashed curves cross.}
\label{fig:V''FixedPoints}
\end{figure}

\begin{figure}[ht]
  \begin{center}
  $
   \begin{array}{ccccc}
     \includegraphics[width=0.3\textwidth,height=0.225\textheight]{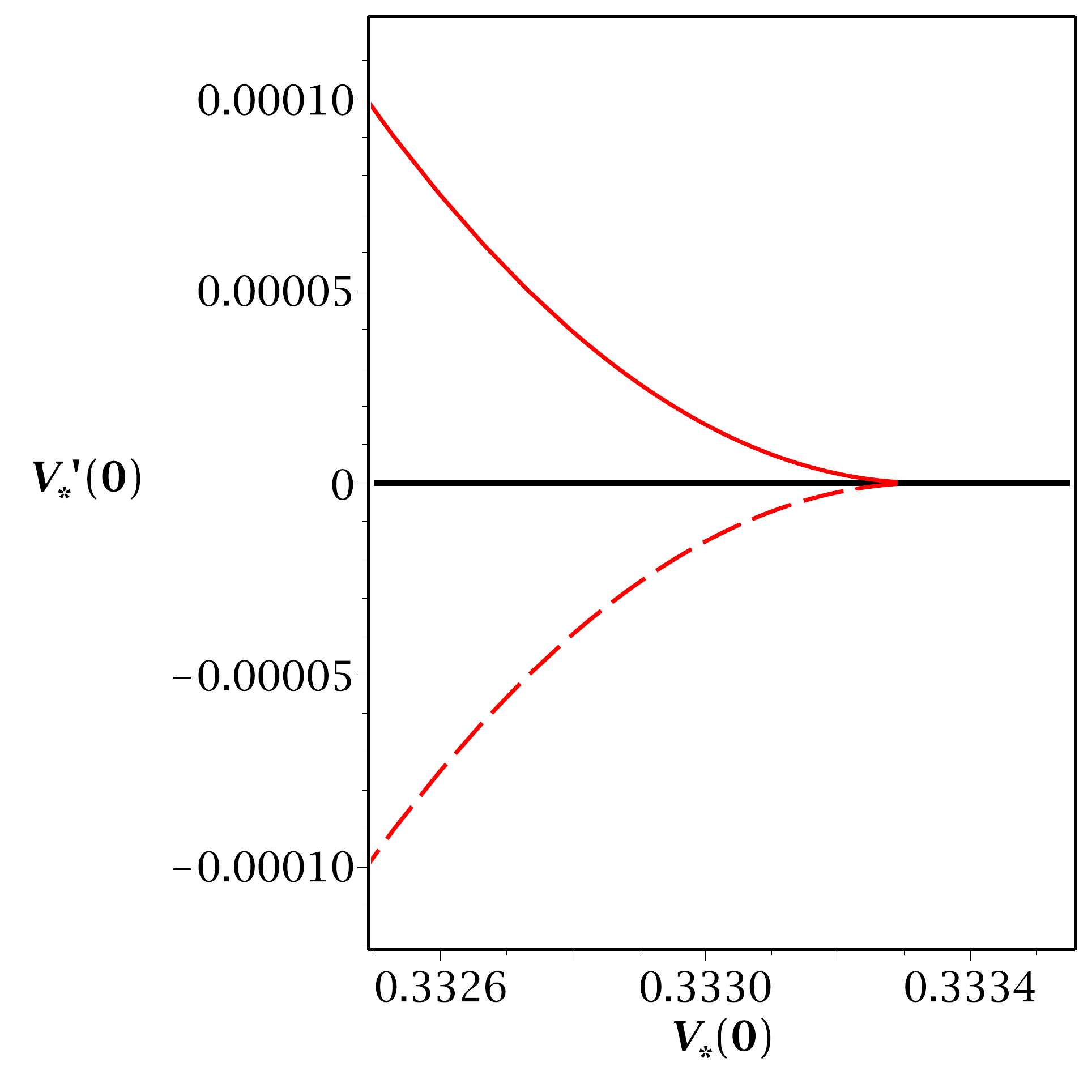}  & &
     \includegraphics[width=0.3\textwidth,height=0.225\textheight]{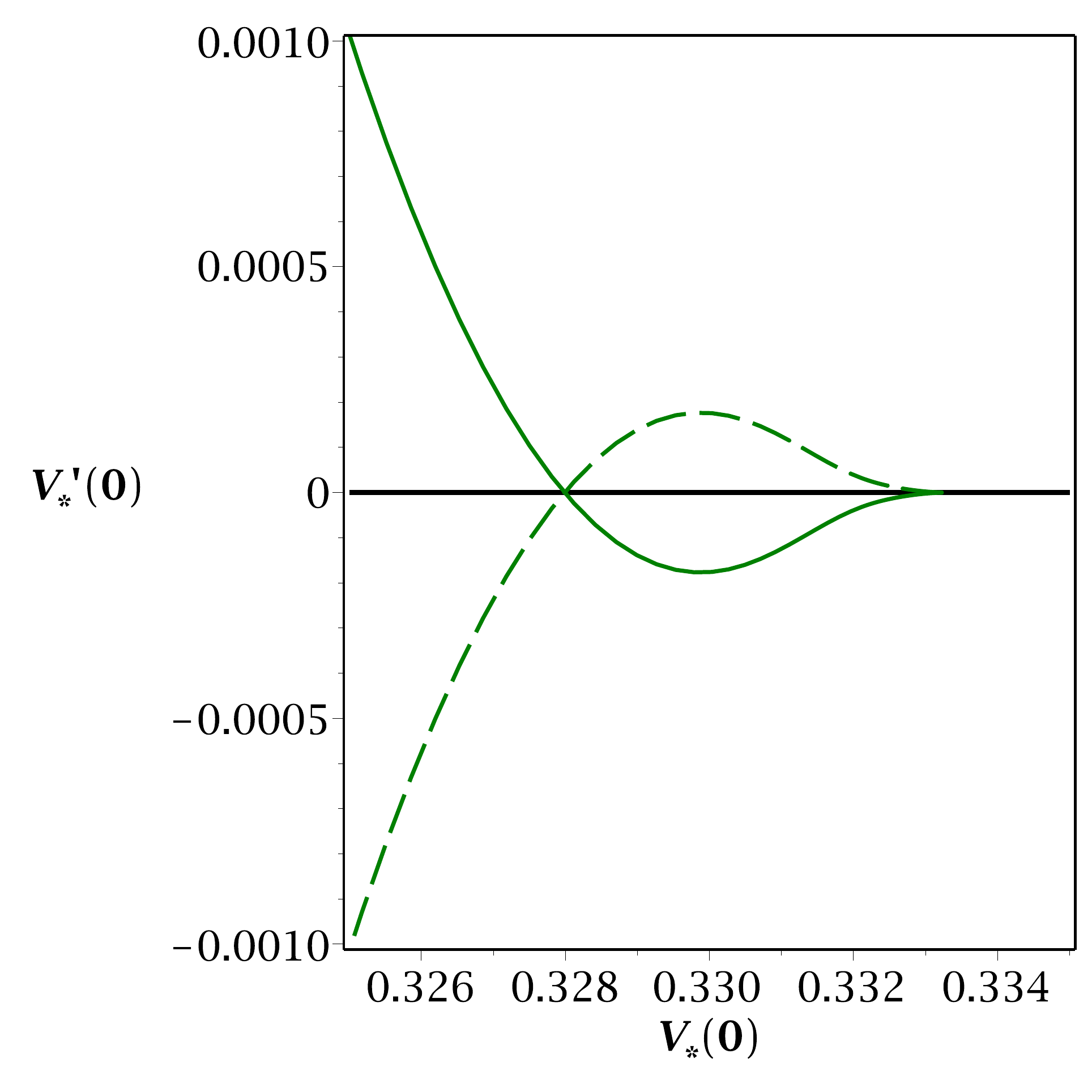} && 
     \includegraphics[width=0.3\textwidth,height=0.225\textheight]{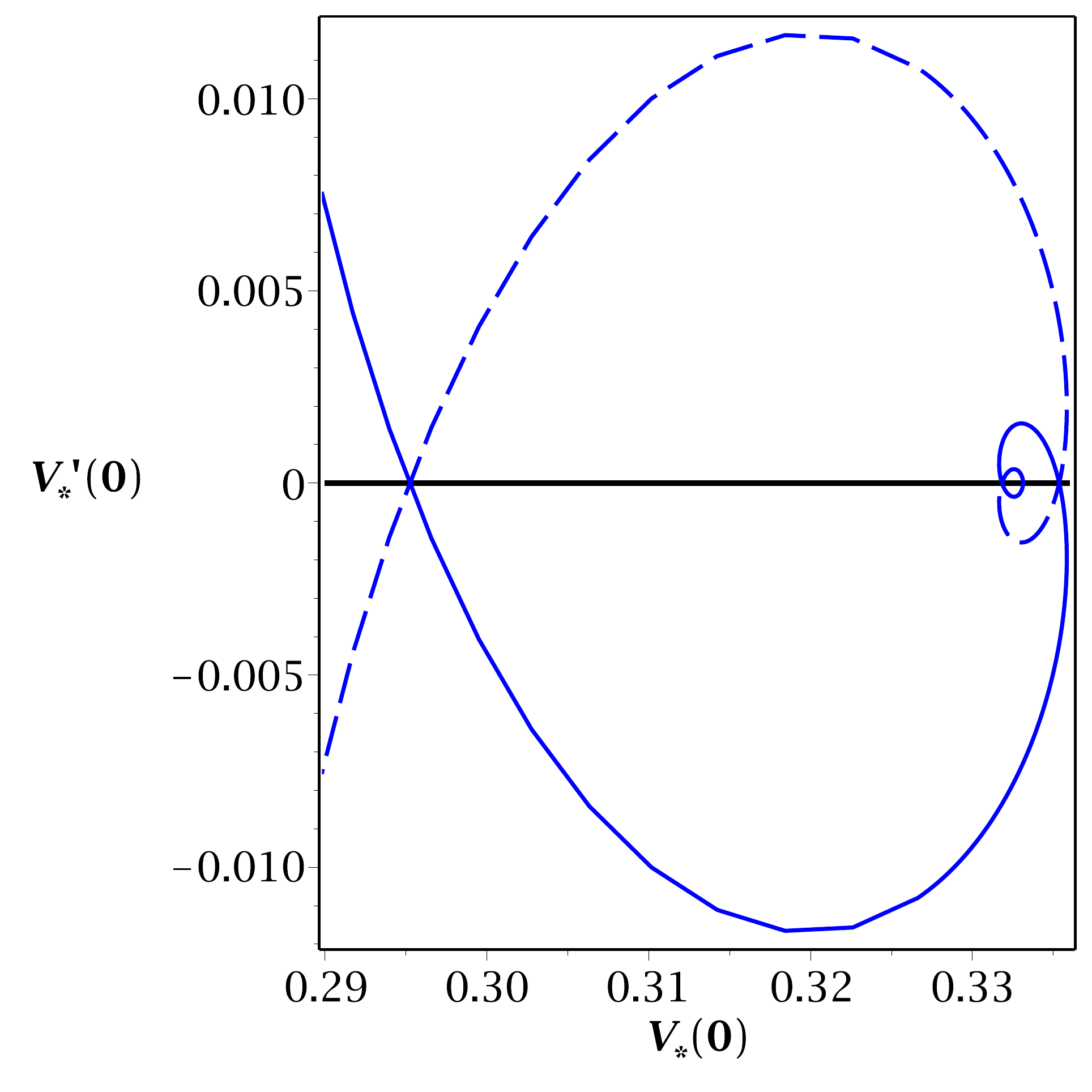}
   \end{array}$
  \end{center}
   \caption{Low values of asymptotic parameter $A$ detail from figure \ref{fig:V''FixedPoints}. Each plot corresponds to the equivalent values of \ensuremath{\al} in figure \ref{fig:V''FixedPoints}. For large \ensuremath{\al} more fixed points appear near the Gaussian fixed point at \ensuremath{V_{*}(0)=1/3}.}
\label{fig:V''Zoom}
\end{figure}

The results of our analysis are shown in fig. \ref{fig:V''FixedPoints} and \ref{fig:V''Zoom} for the cases \ensuremath{\al=0.5,\ 1 \text{ and } 2}. 
If we first discuss the three plots of the full parameter space of \ensuremath{0<A<\infty} for
\ensuremath{V_{*}(0) \text{ vs. } V'_{*}(0)} for \ensuremath{\al=0.5,1\ \&\ 2} in fig. \ref{fig:V''FixedPoints},
we find that for all values of \ensuremath{\al} there
exist at least two fixed point solutions.  As \ensuremath{A\to 0} we get the Gaussian fixed point at \ensuremath{V_{*}=1/3}
that we would expect given the form of the fixed point equation \eqref{equ:V''Fixed}.
Furthermore, for all three values of $\al$ there is a second fixed point solution characterised by $V_*(0)>0.4$
which we identify with the Wilson-Fisher fixed point in the present context. This identification is made by analogy
in shape of the spiral in fig. \ref{fig:V''FixedPoints} to fig. \ref{fig:WF} (left).
As \ensuremath{A\to \infty}
there appears to be another fixed point solution at \ensuremath{V_{*}(0)=V'_{*}(0)=0}, however this point does not have sensible asymptotic behaviour.

From fig. \ref{fig:V''FixedPoints} (right) we can already see that depending on $\al$ there can be more than these two standard fixed points.
This is confirmed by the plots in fig. \ref{fig:V''Zoom} which show magnifications corresponding to the plots
in fig. \ref{fig:V''FixedPoints}.
As we vary \ensuremath{\al} we find that for \ensuremath{\al\gtrapprox 0.85} a first additional non-trivial fixed point
appears which, as can be seen from fig. \ref{fig:V''Zoom} (middle), is still present in the important case $\al=1$ considered in \cite{Litim:2002hj}.
Increasing $\al$ to above one results in even more fixed points, \eg for $\al =2$ there are three additional fixed points,
\cf fig. \ref{fig:V''Zoom} (right). The number of these non-trivial fixed points that appear near \ensuremath{A= 0} increases dramatically as we increase \ensuremath{\al}, \eg for \ensuremath{\al=4} there are in excess of 20 new fixed points.
As we increase \ensuremath{\al} the spiral in \ensuremath{V_{*},V'_{*}} space flattens towards the \ensuremath{V_{*}} axis
and each non-trivial fixed point moves away from the Gaussian fixed point on the \ensuremath{V_{*}} axis.

Since we know that the correct description of a single scalar field in three dimensions includes only  
one non-trivial fixed point we conclude already at this stage that the fixed point equation \eqref{equ:V''Fixed} 
is invalidated by the existence of these additional fixed points.

This conclusion is further supported by an investigation of the eigenspectra of the fixed point solutions described
in this section. Interestingly, creating a plot analogous to fig. \ref{fig:vps} for the non-trivial fixed point solutions in fig.
\ref{fig:V''Zoom} shows that each of these displays a minimum at non-zero field, \ie a decisive obstruction
to redundant operators as expressed in \eqref{equ:redcond}. However, as we will see now, we encounter a very different problem.

Linearising the flow equation \eqref{equ:V''Flow} and assuming separation of variables as in \eqref{equ:perts}
leads to a linear differential equation of third order for the eigenperturbations $v(\phi)$. Since the step function
from the right hand side of the fixed point equation \eqref{equ:V''Fixed} appears in the same form on the right hand
side of the eigenoperator equation, the asymptotic solution $v(\phi)=\phi^{6-2\lambda}$, given by solving its left hand side,
has to be matched onto the solution of the full equation at the same field value $\phi_c=(30A\al)^{-1/4}$ as for the
fixed point equation. Analogous to \eqref{equ:TrunTaylor}, we can then set up an expansion of $v(\phi)$ valid
to the left of the matching point which by matching derivatives up to the second, takes the following form:
\begin{equation*}
 v(u) = (30\al A)^{\frac{1}{2}(\lambda-3)} + 2(\lambda-3)(30\al A)^{\frac{1}{4}(2\lambda-5)} u + (2\lambda-5)(\lambda-3)(30\al A)^{\frac{1}{2}(\lambda-2)} u^2 + \tilde F(u).
\end{equation*}
As in \eqref{equ:TrunTaylor}, we use $\phi = \phi_c -u$ and $u\geq0$ as we are interested in expanding
to the left of the matching point $\phi_c$ and $\tilde F(u)$ is again a remainder that has to vanish faster
than $u^2$ in order to not affect the required matching of $v$, $v'$ and $v''$ to the asymptotic solution at $u=0$.
Writing, \cf \eqref{equ:F(u)gammaf},
\begin{equation*}
 \tilde F(u) = u^\delta \tilde f(u),
\end{equation*}
this translates into the condition $\delta > 2$. It can then be shown using this ansatz in the eigenoperator equation
that its left and right hand sides can only be made equal at the matching point $\phi=\phi_c$ if $\lambda=3$ or $5/2$.
The first eigenvalue corresponds to the vacuum energy, and the second to the odd eigenoperator $v(\phi)=\phi$; 
the eigenoperator equation admits no other global solutions. In particular, apparently none of the non-trivial fixed point solutions
found above possess a relevant eigenoperator resembling the one needed for the Wilson-Fisher fixed point.

We note that these conclusions are valid for the non-trivial fixed points of this section. At the Gaussian 
fixed point $V_*=1/3$ the right hand side of the eigenoperator equation does no longer require a matching procedure
since the step function is identically equal to one. An analysis of the resulting eigenoperator
equation confirms the standard spectrum of relevant eigenoperators of the Gaussian fixed point, \cf below
\eqref{equ:WFeops}.

We therefore see that the flow equation \eqref{equ:V''Flow} gives qualitatively very different results  from both the correct description of fixed points in $d=3$ scalar field theory in sec. \ref{sec:WF} and the 
incorrect description following from the flow equation \eqref{equ:flowLPA} with \eqref{cutoff1} of sec. \ref{sec:gencutoff}.
Nevertheless, we will show in sec. \ref{sec:sWI-Vpp} how the cutoff choice \eqref{cutoff2} in \eqref{equ:cutoffh}
leads to the correct description of single component scalar field theory in $d=3$
with the help of the modified shift Ward identity.

\section{Including the modified shift Ward identity} \label{sec:backgroundfield}
In the previous sections we have seen how the use of cutoffs of form \eqref{equ:cutoffh} with
\eqref{cutoff1} or \eqref{cutoff2} and $\al\neq0$ 
together with the single field approximation has led to a rather severe distortion of the established fixed point structure
for the theory of a single scalar field in $d=3$ dimensions. As we show now, it is however possible to recover the correct results as presented in sec. \ref{sec:WF} by appropriately taking into account the dependence
of the effective average action on the background field through the modified shift Ward identity.

\subsection{Derivation and interpretation} \label{sec:sWI-derivation}
In order to derive the modified shift Ward identity, we start with the partition function
\begin{equation} \label{partfunc}
 Z[J,\pb]=\int \mathcal{D}\vp \exp\left(-S[\vp+\pb]-S_k[\vp,\pb]+ J\cdot \vp\right)
\end{equation}
where we have used the notation
\begin{equation*}
 J\cdot\vp \equiv J^a\vp_a \equiv \int_x J(x)\vp(x),
\end{equation*}
the index $a$ replacing the coordinates $x$. The cutoff action $S_k$ is then written as $S_k[\vp,\pb] = \frac{1}{2}\vp \cdot R_k[\pb] \cdot\vp $.
We note that the classical action $S$ is invariant under the combined shift 
\begin{equation} 
 \label{shifts}
 \pb \mapsto \pb + \eps(x) \qquad \text{and} \qquad \vp \mapsto \vp - \eps(x),
\end{equation}
whereas the cutoff action and the source term break this invariance. Hence we can write
\begin{equation*}
 Z[J,\pb+\eps]=\int \mathcal{D}\vp \exp\left(-S[\vp+\pb]-S_k[\vp-\eps,\pb+\eps]+ J \cdot (\vp-\eps)\right)
\end{equation*}
and, taking $\eps$ to be infinitesimal, the resulting variation of the functional $W=\ln Z$ becomes 
\begin{equation*}
 \frac{\delta W}{\delta \pb}\cdot\eps = \eps\cdot R_k \cdot \frac{\delta W}{\delta J}
 -\frac{1}{2}\frac{\delta W}{\delta J} \cdot \left(\frac{\delta R_k}{\delta \pb_a}\, \eps_a\right) \cdot \frac{\delta W}{\delta J}
 -\frac{1}{2} \mathrm{Tr} \left [\left(\frac{\delta R_k}{\delta \pb_a}\, \eps_a \right) \cdot \frac{\delta^2 W}{\delta J \delta J}\right] - J \cdot \eps,
\end{equation*} where again the dot indicates the contraction of the corresponding free indices.
We now express this equation in terms of the effective average action
\begin{equation} \label{Legendre-trafo}
 \Gamma_k[\vp^c,\pb] = J \cdot \vp^c - W[J,\pb] - S_k[\vp^c,\pb]
\end{equation}
by taking into account the identities
\begin{equation*}
 \vp^c = \frac{\delta W}{\delta J}, \qquad \frac{\delta \Gamma_k}{\delta \pb}=-\frac{\delta W}{\delta \pb}-\frac{\delta S_k}{\delta \pb},
 \qquad \frac{\delta^2 W}{\delta J \delta J} = \left[\frac{\delta^2 \Gamma_k}{\delta \vp^c \delta \vp^c}+R_k\right]^{-1}
\end{equation*}
and arrive at
\begin{equation*}
 \left(\frac{\delta \Gamma_k}{\delta \pb}-\frac{\delta \Gamma_k}{\delta \vp^c}\right)\cdot \eps
 = \frac{1}{2} \mathrm{Tr} \left [\left(\frac{\delta^2 \Gamma_k}{\delta \vp^c \delta \vp^c}+R_k\right)^{-1}\cdot \frac{\delta R_k}{\delta \pb_a}\, \eps_a\right].
\end{equation*}
Since $\eps$ is arbitrary, this identity still holds after dropping it:
\begin{equation}
 \label{sWI}
  \frac{\delta \Gamma_k}{\delta \pb_a}-\frac{\delta \Gamma_k}{\delta \vp^c_a} = \frac{1}{2} \mathrm{Tr} \left [\left( \frac{\delta^2 \Gamma_k}
 {\delta \vp^c \delta \vp^c} + R_k \right)^{-1} \frac{\delta R_k}{\delta \pb_a}\right].
\end{equation}
This is the (modified) shift Ward identity (sWI) capturing the dependence of the effective average action
on the background field as
introduced by the cutoff action breaking the symmetry \eqref{shifts}
of the classical action.

In the context of scalar field theory the sWI has appeared previously in \cite{Litim:2002hj},
expressed in terms of $\tilde \Gamma_k[\phi^c,\pb] = \Gamma_k[\phi^c-\pb,\pb]$ (see footnote \ref{Reuter-vars}).
An analogous sWI for Yang-Mills theory
can be found in \cite{Reuter:1993kw,Litim:1998nf,Reuter:1997gx,Litim:2002ce} and for scalar QED in \cite{Reuter:1994sg}.
An appropriate version of the sWI for conformally reduced gravity has been used in \cite{Manrique:2009uh}.

The structure of the sWI is very close to that 
of the flow equation \eqref{equ:FRGE}, in particular note that the trace is over the same indices, which is a result
of the following alternative way of deriving it. Taking the partition function \eqref{partfunc}
and rewriting it as
\begin{equation*}
 Z[J,\pb]=\int \mathcal{D}\phi \exp\left(-S[\phi]-S_k[\phi-\pb,\pb]+ J\cdot (\phi-\pb)\right),
\end{equation*}
where $\phi=\vp+\pb$ is the total field, 
it suffices to take a functional derivative with respect to 
the background field and follow a similar procedure as in the derivation of the flow equation
\eqref{equ:FRGE} to arrive at the sWI \eqref{sWI}. This more direct derivation
is the one followed in refs. \cite{Reuter:1993kw,Litim:2002ce,Reuter:1994sg,Reuter:1997gx,Manrique:2009uh}.
Through the derivation above however, we find the interpretation as a modified Ward identity
expressing the breaking of the shift symmetry \eqref{shifts}. 

In this context it is important to stress that the sWI \eqref{sWI} represents an additional
constraint on the effective average action and thus needs to be investigated in its own right.
If $\Gamma_k[\vp^c,\pb]$ is a solution of the flow equation \eqref{equ:FRGE} at the exact level and we modify
the effective average action by an arbitrary $k$-independent  functional depending only on the background field,
$\Gamma'_k[\vp^c,\pb]=\Gamma_k[\vp^c,\pb]+F[\pb]$, we obtain a new effective average action which is still
a solution of the flow equation. On the other hand,
if $\Gamma_k[\vp^c,\pb]$ satisfies the sWI \eqref{sWI} the same will no longer be true for $\Gamma'_k[\vp^c,\pb]$, except in the case where $F$ is just a constant.

The reason why the sWI no longer holds for $\Gamma'_k$ is that by virtue of \eqref{Legendre-trafo} we see that 
it corresponds to a different bare action than the one that gives $\Gamma_k$.
If the latter is associated
to the partition function \eqref{partfunc} the former corresponds to the same partition function with the
replacement $S[\vp+\pb] \rightarrow S[\vp+\pb] - F[\pb]$. According to the derivation of the sWI above,
this modification of the path integral would lead to a sWI different from \eqref{sWI} due to the fact
that the bare action in this path integral does no longer respect the shift symmetry \eqref{shifts}.
In other words, the flow equation \eqref{equ:FRGE} does not include the information that in the path integral
it is derived from a bare action $S[\vp+\pb]$ that depends only on the total field instead of on both fields separately.

Indeed, treating the general case now, we note that using a bare action $S[\vp,\pb]$ depending separately on the 
fluctuation field and the background field in the partition function \eqref{partfunc}
will lead to the same flow equation \eqref{equ:FRGE} whereas the sWI \eqref{sWI} would take a very different form.
It is therefore the sWI that contains this crucial bit of information that the
bare action is not a functional of two separate arguments but depends on the total field only.

We thus see that as soon as the background field method is employed and the effective
average action is a functional of both fields separately, the flow equation \eqref{equ:FRGE} will generically
admit multiple solutions as it alone is not enough to determine the background field dependence of the effective average action.

Even though we cannot infer that the sWI \eqref{sWI} is satisfied if the flow equation \eqref{equ:FRGE}
holds, it suffices to satisfy the sWI at one particular point $k=k_0$ along the flow to ensure
that it holds for any scale $k$. If we denote the right hand side of \eqref{Legendre-trafo} by 
$\mathcal{F}[\vp^c,\pb,k]$ the flow equation is just 
$\partial_t \Gamma_k[\vp^c,\pb] = \partial_t \mathcal{F}[\vp^c,\pb,k]$ and similarly the sWI becomes
$\frac{\delta}{\delta \pb}\left\{\Gamma_k[\vp^c,\pb] - \mathcal{F}[\vp^c,\pb,k]\right\}=0$.
Given that the flow equation is satisfied, we see that the time derivative of the sWI
automatically vanishes.	

A second reason for imposing the sWI on the effective average action independently of the flow equation \eqref{equ:FRGE} has its roots in the fact that the separate background field dependence of
$\Gamma_k[\vp^c,\pb]$ comes from the cutoff operator $R_k[\pb]$. A priori, there is a great deal of freedom
in the choice of cutoff operator as far as its dependence on the background field is concerned,
and each choice might lead to a different set of solutions to the flow equation. The sWI \eqref{sWI}
however captures the dependence of the effective average action on the background field
as introduced by the cutoff operator and thus has the potential of re-adjusting the structure
of the solution space of the flow equation accordingly. In sec. \ref{sec:sWI-LPA} we will see 
an explicit example of this.

So far we have assumed that the effective average action
has not been truncated to lie in a subspace of the full theory space. In practice
this is however always necessary, in which case $\Gamma_k[\vp^c,\pb]$ only contains operators
of a certain specified type. In such a case the arguments given above, showing how
the sWI enforces that the bare action depends only on the total field $\vp+\pb$, are no longer
applicable, as we cannot make the connection back to a partition function that gives rise
to the truncated effective average action. In the same way, we cannot generally prove for truncations
that the sWI will be satisfied along the whole flow given that it holds at one particular point of it.
Nevertheless, judging from the discussion above it would seem that for a truncation which is sufficiently
``close'' to an exact solution of the flow equation, explicitly imposing the sWI will increase
the reliability of the truncation.

\subsection{The LPA with the modified shift Ward identity}
\label{sec:sWI-LPA}
As argued in the previous section, the sWI \eqref{sWI} has to be considered in conjunction with
the flow equation \eqref{equ:FRGE}. By following this procedure we will now see how
imposing the sWI in addition to the flow equation allows us to overcome the difficulties encountered
in sec. \ref{sec:LPA} in the presence of a field dependent cutoff operator and, in particular, recover
the standard fixed point structure of 3-dimensional single component scalar field theory.

At the level of the LPA the ansatz we make for the effective average action is
\begin{equation} \label{effact2}
 \Gamma_k[\vp,\pb] = \int_x \left\{ \frac{1}{2}\left(\partial_\mu \vp\right)^2 \right.
 \left. +\frac{1}{2}\left(\partial_\mu \pb\right)^2 + \gamma\partial_\mu\vp \partial^\mu\pb + V(\vp,\pb)\right\},
\end{equation}
where we have canonically normalised the first two derivative terms, included a relative
normalisation constant $\gamma$ in the third term and re-named $\vp^c \rightarrow \vp$.
This truncation represents the generalised LPA in the context of an effective average action
which separately depends on the background field. If we wish to impose $Z_2$-symmetry on
one or separately on both
of the arguments of the effective average action the constant $\gamma$ in this ansatz
would have to be set to zero.

In what follows we will use the general form \eqref{equ:cutoffh} of the cutoff operator,
which we reproduce here for convenience:
\begin{equation} \label{equ:cutoffh1}
 R_k\left(-\partial^2,\pb\right) =\left(k^2+\partial^2-h(\pb)\right) \theta \! \left(k^2+\partial^2-h(\pb)\right),
\end{equation}
We recall that $h$ is an arbitrary function of the background field and renormalisation group time.
Even though we will keep the discussion at the level of a general $h$, we remark that 
the reader is allowed to specialise to either \eqref{cutoff1} or the appropriate generalisation of \eqref{cutoff2} at any stage.
Inserting this and \eqref{effact2} into the sWI \eqref{sWI}, performing the trace and 
adopting dimensionless variables after absorbing a constant into the definition of the 
potential and the fields, we are led to the sWI in the LPA which is the first line
in the following system of equations. A similar calculation for the flow equation \eqref{equ:FRGE}
leads to the second line:
\begin{subequations}
\label{system}
\begin{align}
 \partial_\vp V-\partial_{\pb} V &= \frac{h'}{2} \frac{(1-h)^{d/2}}{1-h+\partial^2_{\vp}V}\,\theta(1-h), \label{sWI-LPA} \\
  \partial_t V - \frac{1}{2}(d-2)\left(\vp \partial_\vp V +\pb \partial_{\pb}V\right) +dV 
 &= \frac{(1-h)^{d/2}}{1-h+\partial^2_{\vp}V} \left(1-h-\frac{1}{2}\partial_t h +\frac{1}{4}(d-2)\pb h'\right)\theta(1-h). \label{flow2}
\end{align}
\end{subequations}
In these equations we have re-instated the space dimension $d$.
We thus obtain a coupled system of non-linear partial differential equations governing the dependence of
the potential $V$ on the background field and the classical field.
It is instructive to consider the case \eqref{cutoff1} and explicitly set $h(\pb)=\al \pb^2$, $d=3$ in \eqref{flow2}.
Restricting
ourselves to fixed points, the result is 
\begin{equation}
 \label{flow2-special}
 3V_* -\frac{1}{2}\left(\vp \partial_\vp V_* +\pb \partial_{\pb}V_*\right) 
  =\frac{\left(1-\al \pb^2\right)^{3/2}}{1-\al \pb^2 +\partial_\vp^2 V_*}\left(1-\frac{1}{2}\al\pb^2\right)\theta\left(1-\al\pb^2\right),
\end{equation}
which we compare to the corresponding fixed point equation \eqref{equ:FPgen} obtained in the single field approximation.
As described below
\eqref{approx} this approximation is obtained by neglecting the remainder $\hat \Gamma_k$ in 
\eqref{approx} from the outset and identifying the background field with the total field
$\pb = \phi$ after the evaluation of the Hessian in \eqref{equ:FRGE}.
If we make this identification in \eqref{flow2-special}, which also implies $\vp=0$,
we recover \eqref{equ:FPgen}. 

Being able to contrast \eqref{flow2-special} with its approximated version \eqref{equ:FPgen} makes the 
consequences of neglecting $\hat\Gamma_k$ in \eqref{approx} stand out. The right hand side of \eqref{flow2-special}
depends on the fluctuation field only through the second derivative of the potential
whereas on the right hand side of the approximated equation \eqref{equ:FPgen} every single term 
contributes. It is therefore to be expected that the solutions to the two equations will show
strong qualitative differences and we will confirm this in the following.

We first come back to an analysis of the full coupled system \eqref{system}.
In order to gain insight into the combined solution space we consider the following change of variables,
\begin{equation} \label{changevars}
 V = (1-h)^{d/2}\hat V, \qquad \vp = (1-h)^\frac{d-2}{4} \hat \vp -\pb, \qquad t = \hat t -\ln \sqrt{1-h}
\end{equation}
in the region $h(\pb)<1$.
In terms of the running scale $k$, the last equality translates into $\hat k =k\sqrt{1-h}$ which
is precisely the replacement that
is needed in \eqref{equ:cutoffh1} in order to transform this cutoff operator into the standard field independent
cutoff operator where $h=0$.\footnote{To see this, we should keep in mind that $[h]=2$ in \eqref{equ:cutoffh1} whereas
in \eqref{changevars} we use the rescaled version of $h$ with $[h]=0$.} 
Since we are working with dimensionless variables, this transformation of the renormalisation group scale
$k$ induces corresponding transformations on the field and the potential as expressed in the first two equations
in \eqref{changevars}.

Applying the change of variables \eqref{changevars} to \eqref{system} leads to the equivalent system
\begin{subequations}
\label{systemt}
\begin{align}
 \partial_{\pb} \Vh &= \frac{h'}{2(1-h)}\, \mathcal{W}, \label{sWI-LPAt} \\
  \frac{1}{2}(d-2)\pb \partial_{\pb}\Vh &= \frac{1}{1-h}\left(1-h-\frac{1}{2}\partial_t h +\frac{1}{4}(d-2)\pb h'\right)  \mathcal{W}, \label{flow2t}
\end{align}
\end{subequations}
where we have abbreviated
\begin{equation*} 
 \mathcal{W}=\partial_{\hat t} \hat V+ d\hat V-\frac{1}{2}(d-2)\hat\vp \partial_{\hat \vp} \hat V - \frac{1}{1+\partial^2_{\hat \vp} \hat V}.
\end{equation*}
By combining the two equations in \eqref{systemt}, we see immediately that this system is in turn equivalent to
\begin{equation} \label{systemfinal}
\partial_{\pb} \Vh 	=0, \qquad \mathcal{W}			=0
\end{equation}
 as long as $h$ does not satisfy $1-h-\frac{1}{2}\partial_t h =0$.
However, the last equation has the solution $h(\pb)=1+c(\pb)\exp(-2t)$, where $c$ is an arbitrary function of the
background field. If we substitute this form of $h$ into the cutoff operator \eqref{equ:cutoffh1} we find that it
is actually independent of the scale $k$, \ie it does not lead to a renormalisation group flow and thus we can
safely discard this special case.

From now on we implement the condition $\partial_{\pb} \Vh=0$ in \eqref{systemfinal}
by dropping the dependence of $\Vh$ on the background field
and tacitly taking $\Vh$ to depend only on the transformed field $\ph$ and renormalisation group time $\that$.
We can therefore conclude that the original system \eqref{system} is fully equivalent to $\mathcal{W}=0$, \ie
\begin{equation} \label{reduced}
 \partial_{\hat t} \hat V+ d\hat V-\frac{1}{2}(d-2)\hat\vp \partial_{\hat \vp} \hat V = \frac{1}{1+\partial^2_{\hat \vp} \hat V},
\end{equation}
in the sense that the change of variables \eqref{changevars} supplies a one to one mapping 
between the solutions $V_t(\vp,\pb)$ of \eqref{system} and the background field independent
solutions $\hat V_{\hat t}(\hat \vp)$ of the reduced flow equation \eqref{reduced}.
This reduced flow equation however is precisely the flow equation we would 
have obtained with a standard background field independent cutoff operator, given
by \eqref{equ:flowLPA} with $h=0$. 

As we can see from the simplified system \eqref{systemt},
it is only through the requirement that the sWI \eqref{sWI-LPAt} hold
that the solutions of \eqref{reduced} are actually independent of the background field.
This is the additional crucial ingredient supplied by the sWI which is not contained in 
the flow equation \eqref{flow2t} alone.

The possibility to reduce the renormalisation group flow \eqref{flow2} for any cutoff operator
of form \eqref{equ:cutoffh1} to the standard flow \eqref{reduced} is a striking manifestation
of universality. Even though the flows \eqref{flow2} may potentially look very different from
the standard flow \eqref{reduced} depending on the choice of $h$, exploiting the sWI paves the way
to recognising their equivalence.

In order to flesh out the details of this equivalence of renormalisation group flows, let us specialise
to the investigation of their fixed points. At a fixed point we drop all $t$-dependence
in the flow equation \eqref{flow2} and thus the dimensionless potential
has to be renormalisation group time independent, $\partial_t V=0$. Accordingly, we also require $\partial_t h=0$ at a fixed point
and denote the corresponding $t$-independent function by $h_*$.
Using \eqref{changevars} and the chain rule, we see that setting $\partial_t V =0$ and $\partial_th =0$
in \eqref{flow2} is equivalent to setting $\partial_{\hat t} \hat V=0$ in the reduced flow \eqref{reduced}.
As expected, we therefore find a one to one mapping of the fixed points of the two renormalisation group flows
which is given by the fixed point version of the change of variables \eqref{changevars}, \ie
\begin{equation}
 \label{FPs-map}
 V_*(\vp,\pb)=(1-h_*(\pb))^{d/2}\,\hat V_*\!\!\left((1-h_*(\pb))^{\frac{2-d}{4}}(\vp+\pb)\right).
\end{equation}
It now becomes apparent how the inaccurate fixed point structure of the renormalisation group flow
in sec. \ref{sec:gencutoff} is avoided by the approach described in this section. Specialising
the fixed point version of \eqref{reduced} to $d=3$ we recover the fixed point equation \eqref{equ:fpWF},
accurately describing the fixed point structure of 3-dimensional single component scalar field theory.
Through \eqref{FPs-map} we conclude that even the full background field dependent flow \eqref{flow2}
admits only the Gaussian and Wilson-Fisher fixed points originating from \eqref{equ:fpWF},
provided we also impose the pertaining sWI \eqref{sWI-LPA}.
A similar resolution of the inaccurate fixed point structure of sec. \ref{sec:LPALitim} is only partly
achieved by the map \eqref{FPs-map} as in this case this relation between fixed point solutions 
still takes the form of a differential equation. We will come back to this and
complete the equivalence in sec. \ref{sec:sWI-Vpp}.

Taking it one step further, the equivalence of \eqref{system} to \eqref{reduced}
also manifests itself at the level of the eigenspectra and eigenoperators of these flows.
Proceeding in the usual way, we write
\begin{equation}\label{linVt}
 V_t(\vp,\pb)=V_*(\vp,\pb) + \eps\, v(\vp,\pb)  \exp(-\lambda t)
\end{equation}
and require that the original system \eqref{system} be satisfied up to linear order in $\eps$.
Similarly, for
\begin{equation} \label{linVht}
 \hat V_{\hat t}(\hat \vp) = \hat V_*(\hat \vp) + \eps \, \hat v(\hat \vp) \exp(-\lambda \hat t)
\end{equation}
we impose that \eqref{reduced} hold up to linear order in $\eps$. Our goal is to show by using
the correspondence \eqref{changevars} that for every eigenoperator $v$ with associated eigenvalue
$\lambda$ we find a corresponding eigenoperator solution $\vh$ of the linearisation of \eqref{reduced}
with the same eigenvalue $\lambda$ and vice versa, thereby proving equality of eigenspectra.
In order to linearise the flow equation \eqref{flow2} after substituting \eqref{linVt} we
also need to perturb the function $h$ according to
\begin{equation}
\label{linh1}
 h_t(\pb) = h_*(\pb) + \eps \, \delta h(t,\pb).
\end{equation}
With this and \eqref{linVt}, the change of variables \eqref{changevars} becomes at order $\eps$,
\begin{equation} \label{dh-tdep}
 v \exp(-\lambda t) = (1-h_*)^\frac{d-\lambda}{2}\vh \exp(-\lambda t)
		      -\frac{\delta h}{2}\frac{(1-h_*)^{\frac{d}{2}-1}}{1+\partial_{\ph}^2\Vh_*}, 
\end{equation}
where the variables are now related by the fixed point version of the change of variables \eqref{changevars},
\begin{equation}
\label{linchangevars}
 V_* = (1-h_*)^{d/2}\hat V_*, \qquad \vp = (1-h_*)^\frac{d-2}{4} \hat \vp -\pb, \qquad t = \hat t -\ln \sqrt{1-h_*}.
\end{equation}
We then recognise from \eqref{dh-tdep} that $\delta h \exp(\lambda t)$ has to be $t$-independent and thus
we can write 
\begin{equation}
 \label{linh2}
   h_t(\pb) = h_*(\pb) + \eps \, \kappa(\pb)\exp(-\lambda t).
\end{equation}
This separable form of the perturbation $\delta h$
in \eqref{linh1} is a necessary requirement if the two systems \eqref{system} and \eqref{reduced} are to possess
the same eigenspectra and is tied to the separability used in \eqref{linVt} and \eqref{linVht}.
In particular, it is applicable if the function $h$ depends on renormalisation group time $t$
only through the potential $V$ as in \eqref{cutoff2}.
Applying this to \eqref{dh-tdep}, we obtain the relation between the original eigenoperator $v(\vp,\pb)$ and
the eigenoperator $\vh(\ph)$,
\begin{equation} \label{rel-eops}
 v = (1-h_*)^\frac{d-\lambda}{2}\vh
		      -\frac{\kappa}{2}\frac{(1-h_*)^{\frac{d}{2}-1}}{1+\partial_{\ph}^2\Vh_*}. \end{equation}
Using \eqref{linVt} and \eqref{linh2} in the original system \eqref{system}, we are led to the following linearised
equations for the eigenoperator $v$:
\begin{subequations}
\label{linsystem}
\begin{align}
 \partial_\vp v - \partial_{\pb} v = \frac{1}{2}\frac{(1-h_*)^{d/2}}{1-h_*+\partial_\vp^2 V_*}\left(\kappa'-\frac{d}{2}
				      \frac{\kappa}{1-h_*}h_*'
				      -\frac{\partial_\vp^2 v-\kappa}{1-h_*+\partial_\vp^2V_*} h_*'\right),
				      \label{sWI-lin} \\
(d-\lambda)v-\frac{1}{2}(d-2)(\vp+\pb)\partial_\vp v = \frac{(1-h_*)^{d/2}}{1-h_*+\partial_\vp^2 V_*}
				      \left((\lambda-d-2)\frac{\kappa}{2}
				      -\frac{(1-h_*)(\partial_\vp^2 v - \kappa)}{1-h_*+\partial_\vp^2 V_*} \right). \label{flow2-lin}
\end{align}
\end{subequations}
The first equation is the linearised sWI and the second equation comes from linearising the flow
equation \eqref{flow2} and subtracting the linearised sWI multiplied by the factor $\frac{1}{2}(d-2)\pb$.

It is then a straightforward calculation to confirm that the linearised system \eqref{linsystem} is fully
equivalent to the linearisation of the reduced flow equation \eqref{reduced} according to \eqref{linVht}, given by
\begin{equation}
 \label{linreduced}
 (\lambda-d)\vh + \frac{1}{2}(d-2)\ph \vh' = \frac{\vh''}{\left(1+ \Vh_*''\right)^2}.
\end{equation}
As expected from the equivalence of the full renormalisation group flows \eqref{system} and \eqref{reduced}
described by the change of variables \eqref{changevars}, the equivalence at the linearised level
is given by the adapted change of variables \eqref{linchangevars}
and the relation between the eigenoperators \eqref{rel-eops}, mapping an eigenoperator solution $v(\vp,\pb)$
of \eqref{linsystem} 
with eigenvalue $\lambda$ onto the eigenoperator solution $\vh(\ph)$ of \eqref{linreduced}
with the same eigenvalue $\lambda$. Hence the eigenspectrum of the flow equation \eqref{flow2}
complemented by the sWI \eqref{sWI-LPA} is identical to the eigenspectrum of the reduced flow
\eqref{reduced}. 

Specialising to $d=3$ in \eqref{linreduced}, we recover the eigenoperator equation \eqref{equ:WFeops}
and thus we conclude that also at the linearised level, the inaccurate description of 3-dimensional
single component scalar field theory described in sec. \ref{sec:gencutoff} and sec. \ref{sec:missingvacua}
can be rectified completely by keeping track of the background field through the sWI \eqref{sWI-LPA}.\footnote{As
for the equivalence of fixed points, the relation \eqref{rel-eops} is still a differential equation in
the case \eqref{cutoff2} and we will take care of this complication in sec. \eqref{sec:sWI-Vpp}.}

We remark that the discussion so far is restricted to the range $h(\pb)<1$ where the change of variables
\eqref{changevars} is valid. For $h(\pb)>1$ the right hand sides of the system \eqref{system} vanish
and we obtain a different solution which then has to be matched onto the previous solution at $h(\pb)=1$.
Since we are mostly interested in $d=3$ dimensions let us briefly illustrate how this matching
can be done for the fixed point and eigenoperator solutions in this case.

Focussing on fixed point solutions first, the solution of \eqref{sWI-LPA} and 
the fixed point version of \eqref{flow2} for $h_*(\pb)>1$ is $V_*(\vp,\pb)=A(\vp+\pb)^6$,
in analogy to the solution of the left hand side of \eqref{equ:FPgen}. This solution needs to
be matched onto the solution \eqref{FPs-map} at $h_*(\pb) \rightarrow 1$ for both possibilities,
the Wilson-Fisher fixed point and the Gaussian fixed point. 
Concentrating on the Wilson-Fisher fixed point, we see that
the argument of $\hat V_*$ in \eqref{FPs-map} diverges in this limit since $d>2$.\footnote{Depending on the precise
form of $h_*$ there could be an exception to this if $\vp+\pb \rightarrow 0$ at the same time. In what follows
we will therefore assume that $\vp \neq -\pb$ in the limit $h_*(\pb)\rightarrow 1$.}
This implies that we can make use
of \eqref{equ:VWFasy} describing the asymptotic behaviour of $\hat V_\mathrm{WF}$ and obtain
\begin{equation} \label{potsolasy}
V_\mathrm{WF}(\vp,\pb) = A_\mathrm{WF}(\vp+\pb)^6 + \frac{(1-h_*)^{5/2}}{150A_\mathrm{WF}(\vp+\pb)^4} + \dots
\end{equation}
in the limit $h_*(\pb)\rightarrow 1$. All subleading terms in this expansion vanish individually 
in the limit we are considering. Assuming that therefore the remainder vanishes as well, we find perfect matching
between the two solutions at $h_*(\pb)=1$. From \eqref{potsolasy} we can also appreciate
that this global solution is smooth in $\vp$ at $h_*(\pb)=1$ and therefore smooth everywhere in $\vp$.
On the other hand, we can see that the third and all higher derivatives of $V_\mathrm{WF}$ with respect 
to the background field $\pb$ will diverge as $h_*(\pb)\rightarrow 1$ as caused by the subleading terms
in \eqref{potsolasy}. This coincides with the observation we made just below \eqref{asyexpcoeffs}
and the same conclusion can also be drawn from the sWI \eqref{sWI-LPA} itself, once we know that $V_\mathrm{WF}$
is smooth in $\vp$. As before, these divergences of derivatives of the fixed point potential
with respect to the background field 
are a direct consequence of the cutoff choice \eqref{equ:cutoffh1}. However we see that as a result of a full treatment
of the background field dependence as dictated by the sWI they do not occur for the classical field $\vp$. 
The same statements are of course true for the Gaussian fixed point whose global form we can write down explicitly:
\begin{equation*}
 V_\mathrm{G}(\vp,\pb)=\frac{1}{3}(1-h_*(\pb))^{3/2} \theta(1-h_*(\pb)).
\end{equation*}
Note that this potential is in fact only a function of the background field.

In order to investigate a similar extension of the eigenoperator solutions $v(\vp,\pb)$ into the range
$h(\pb)>1$, we first note that also at the linearised level the condition $h(\pb)>1$ is equivalent to
$h_*(\pb)>1$ as the derivative of the step function does not lead to a contribution to the linearised equations.
If we want to match the solution \eqref{rel-eops} valid in the range $h_*	(\pb)<1$ 
to the solution $v(\vp,\pb)=|\vp+\pb|^{-2\lambda+6}$ of the left hand sides
of \eqref{linsystem} we can proceed in a similar way as for the fixed point solution.
For the Wilson-Fisher fixed point we can make use of the asymptotic expansion \eqref{equ:WFeopsasy}
of the eigenoperator solutions, the argument of $\hat v$ in \eqref{rel-eops} being divergent in the limit $h_*(\pb)\rightarrow 1$.
With the second term in \eqref{rel-eops} vanishing in this limit, we find that
\begin{equation*}
v(\vp,\pb) = |\vp+\pb|^{-2\lambda+6} -\frac{1}{4500}\frac{(2\lambda-5)(2\lambda-6)}{A_\mathrm{WF}}
\frac{(1-h_*)^{5/2}}{|\vp+\pb|^{2\lambda+4}} + \dots
\end{equation*}
and by the same arguments as below \eqref{potsolasy} we see
that $v$ is a globally defined function which is smooth in $\vp$ and twice differentiable in $\pb$.
As a consequence of the step function in our choice of cutoff all higher derivatives of $v$
with respect to the background field diverge as $h_*(\pb)\rightarrow 1$.

\subsection{Dealing with a cutoff depending on the potential} \label{sec:sWI-Vpp}
 A crucial step in the process of showing how the sWI reduces the background field dependent flow
 \eqref{flow2} to the standard flow of a single scalar field in the LPA \eqref{reduced} is represented
 by the change of variables \eqref{changevars}. This is a well-defined change of variables if the cutoff modification $h$
 is an explicit function of the background field but it does not supply us with a direct definition of $V(\vp,\pb)$
 if $h$ is a function of the potential itself. For example, this is the case for 
 \begin{equation}
 \label{hVpp}
  h_t(\pb)= \left. \partial^2_\vp V_t(\vp,\pb)\right|_{\vp=\pb},
 \end{equation}
where in this instance we have emphasised the $t$-dependence that is usually left implicit .
This is the appropriate generalisation of \eqref{cutoff2} in the context of an effective average action
fully depending on the background and fluctuation fields at the level of the LPA.
We will now show how the steps of the previous section can be adapted for this particular modification
of the cutoff operator \eqref{equ:cutoffh1} in order to illustrate how the sWI again leads to the correct description
of single component scalar field theory in $d=3$, thereby comprehensively rectifying the inaccurate results
found in sec. \ref{sec:LPALitim}.

We start by implicitly defining a new field variable $\tilde \vp$ through
\begin{equation}
 \label{tphi}
 \pb = \frac{1}{2}\frac{\tilde \vp}{\left(1+\Vh''(\tilde \vp)\right)^{1/4}},
\end{equation}
where as before $\Vh$ is a solution of the reduced flow equation \eqref{reduced}. This is trivially possible at the Gaussian fixed point 
but also at the Wilson-Fisher fixed point as plotting the right hand side of \eqref{tphi} proves it to be a strictly monotonically increasing
function. The same will be true for small $t$-dependent perturbations around either of these fixed points and the following steps hold whenever this relation can be inverted 
to obtain $\tilde \vp$ as a function of the background field and renormalisation group time.
In general, the allowed range of $\pb$ in $\tilde \vp(\pb)$ is bounded, \eg at the Wilson-Fisher fixed point 
we have
\begin{equation*}
 \lim_{\tilde \vp \rightarrow -\infty} \pb(\tilde \vp) = -\frac{1}{2\left(30 A_\mathrm{WF}\right)^{1/4}}
 <  \pb(\tilde \vp) < \frac{1}{2\left(30 A_\mathrm{WF}\right)^{1/4}} = \lim_{\tilde \vp \rightarrow \infty} \pb(\tilde \vp)
\end{equation*}
as can be deduced from its asymptotic form \eqref{equ:VWFasy}.
We then define the following change of variables
\begin{equation}
\label{changevarsVpp}
\begin{gathered}
 \ph = \left(1+\Vh''(\tilde \vp)\right)^{1/4} (\vp + \pb), \qquad V(\vp,\pb) = \left(1+\Vh''(\tilde \vp)\right)^{-3/2} \Vh(\ph), \\
  t = \hat t + \ln \sqrt{1+\Vh''(\tilde \vp)},
\end{gathered}
\end{equation}
where we always regard $\tilde \vp$ as a function of $\pb$. A short calculation then shows
\begin{equation}
\label{hVpp-relation}
 1-h(\pb)=1-\left. \partial^2_\vp V(\vp,\pb)\right|_{\vp=\pb}=\frac{1}{1+\Vh''(\tilde \vp)}.
\end{equation}
Using this identity, the change of variables \eqref{changevarsVpp} assumes the form of the change of variables
\eqref{changevars} of the previous section, implying that the results between the equivalence of the complicated
system \eqref{system} with \eqref{hVpp} and the standard flow \eqref{reduced} carry over to the present case.

In particular, the relation \eqref{hVpp-relation} gives 
\begin{equation} \label{solh}
 h_*(\pb) = \left. \partial^2_\vp V_*(\vp,\pb)\right|_{\vp=\pb} = \frac{\Vh_*''(\tilde \vp)}{1+\Vh_*''(\tilde \vp)},
\end{equation}
through which the fixed point solutions are given by \eqref{FPs-map} with $d=3$, $\Vh_*$ being either the Gaussian fixed point
or the Wilson-Fisher fixed point of the standard flow \eqref{reduced}. 
In the case of the Gaussian fixed point we simply have $h_*=0$ and $V_*=1/3$.

Furthermore, \eqref{hVpp-relation} also implies that in \eqref{linh2} we have
\begin{equation*}
\kappa(\pb) =  \left. \partial^2_\vp v(\vp,\pb)\right|_{\vp=\pb} = \frac{\hat v''(\tilde \vp)}{\left(1+
\Vh_*''(\tilde \vp)\right)^2}
\end{equation*}
which together with \eqref{solh} defines the eigenoperator solutions via \eqref{rel-eops} and as an immediate consequence  the eigenspectra
of the standard flow \eqref{reduced} in $d=3$ are again shown to be equal to the ones of the corresponding fixed points of the system \eqref{system}.
For example, at the Gaussian fixed point $V_*=1/3$ the eigenoperator solutions of \eqref{linsystem} are
\begin{equation*}
v(\vp,\pb)=\vh(\vp+\pb)-\frac{1}{2}\vh''(2\pb),
\end{equation*}
where $\vh$ is any eigenoperator solution of \eqref{linreduced} at the Gaussian fixed point $\Vh_*=1/3$.

This completes our discussion of how the inaccurate results of sec. \ref{sec:gencutoff} and \ref{sec:LPALitim}
in the context of the single field approximation with field dependent cutoff operators can be understood and corrected 
by the use of the modified shift Ward identity \eqref{sWI}.

\section{Conclusions} \label{sec:Disc}

Motivated by the use of the background field method in asymptotic safety for gravity, the aim of this work was
to investigate a similar approach in the less involved context of the LPA in scalar field theory.
The background field method opens up a wide variety of possibilities for background field dependent
cutoff operators such as \eqref{equ:cutoffh}. If it is then decided to neglect the ensuing separate
dependence of the effective average action on the background field by adopting the single field approximation
as described below \eqref{approx} we have shown in sec. \ref{sec:LPA} that both quantitatively and qualitatively inaccurate
results follow, even in the context of a single scalar field. 

Fixed points that would be present for field independent cutoffs can disappear, and new fixed points 
accompanied by differing numbers of relevant eigenoperators
can be created, depending on the precise form of background field dependence of the cutoff.
More than that, as we have seen in sec. \ref{sec:missingvacua}, the explicit form of a given fixed point
can be altered in such a way that relevant eigenoperators become redundant and thus are no longer part of the true
space of eigenoperators, even though they would have been for a field independent cutoff.
We have investigated these issues in the context of 
two qualitatively different classes of field
dependent cutoff operators for which we could control the background field dependence
through an additional parameter $\al$. 
Based on the general form \eqref{equ:cutoffh}, the first class consisted of the choice \eqref{cutoff1},
where the background field dependence is explicit. This has to be contrasted with the second class, \eqref{cutoff2},
consisting of cutoff operators with an implicit dependence on the background field as supplied by the potential.
Depending on $\al$, in both cases we arrive at a variety of the pathologies described above. For example for the first class, if $\al$ is positive 
yet not too large, we lose the Gaussian fixed point but retain the Wilson-Fisher fixed point
and even though the eigenvalue for its relevant direction depends on $\al$ it does not turn
into a redundant eigenoperator. For $\alpha\ge1/9$, we already lose also the Wilson-Fisher fixed point.
As soon as $\al$ is negative however,  additional fixed points are created whose odd eigenperturbations
are all redundant due to the lack of a non-trivial vacuum solution.
For the second class of cutoff operators we have seen in sec. \ref{sec:LPALitim} that additional
fixed points are present for $\al\gtrapprox0.85$
and all of them display a non-trivial vacuum solution, but (for any $\alpha>0$) the standard form of eigenoperator 
equation does not admit any non-trivial solutions.
Our results thus point to the expectation
that whatever the precise form of field dependence of the cutoff operator, \emph{some qualitative} 
deviation from the results without field dependent cutoffs has to be anticipated.

Fortunately as we have seen in sec. \ref{sec:backgroundfield}, all these problems related
to the single field approximation within the background field method, 
can be successfully avoided provided we take the full dependence
of the effective average action on the background and the classical field into account,
where its dependence on the background field is governed by the modified shift Ward identity
\eqref{sWI}. We emphasise that the sWI is a direct consequence of following through how the
shift symmetry \eqref{shifts} is broken by the cutoff action and the source term in
the original partition function \eqref{partfunc} of the theory. The sWI has to be regarded
as complementary to the flow equation \eqref{equ:FRGE} itself and a full analysis has 
to incorporate both of them. In fact, as discussed in sec. \ref{sec:sWI-derivation},
at the exact level the flow equation alone is not enough to uniquely determine the background field dependence
of the effective average action and therefore the sWI is required as an additional constraint on the flow.
As such, a specific truncation of the effective average action that satisfies both the flow equation
and the sWI can be regarded as more reliable than if it satisfied only the flow equation.

Even though taking the sWI into account adds a certain degree of difficulty
to the search for fixed points and the investigation of their eigenspectra, the clear way
in which all the issues of sec. \ref{sec:LPA} stemming from field dependent cutoffs have been resolved in
sec. \ref{sec:sWI-LPA} and \ref{sec:sWI-Vpp} constitutes firm evidence that use of the sWI can be very powerful. In the present
case of 3-d, single component scalar field theory in the LPA, the system of sWI 
and the non-perturbative renormalisation group flow \eqref{flow2} with the general class of background field dependent
cutoff operators \eqref{equ:cutoffh1} enabled us to show the full equivalence to the standard flow
with field independent cutoff operator \eqref{reduced}. As shown in sec. \ref{sec:sWI-Vpp}, this method
can also work if the background field dependence in the cutoff is through the potential itself.

These results of equivalence of renormalisation group flows can be seen as the fingerprint of universality.
In the bigger picture consisting of the effective average action depending on the fluctuation and the background
field, as governed by the flow equation and the sWI, the large class of different cutoff operators \eqref{equ:cutoffh1}
lead to 
exactly the same physical data
for the fixed points and eigenspectra. Indeed, they all lead even to the same 
global renormalisation group flow.

Of course we would hope that, once correctly implemented, and before any approximation, we should recover universality with respect to different forms of cutoff, this being Wilson's essential insight  \cite{Wilson:1973}. Generically however we  would also expect this universality to be broken by the truncation scheme. As we have seen, remarkably this does not happen for the LPA. 

In scalar field theory the use of the background field method is artificial and not needed
in the analysis of the flow equation \eqref{equ:FRGE}. For the corresponding flow equation
in the context of Yang-Mills theory or gravity however, the background field method is a
valuable tool in order to realise gauge invariance or diffeomorphism invariance, \cf
\cite{Reuter:2012,Reuter:1993kw}.
Thus, coming back to the original motivation for this article, let us comment on the possible
implications of our results for asymptotic safety. 
The vast majority of studies searching for a non-perturbative fixed point for quantum gravity
have been carried out in the single field approximation and a large amount
of evidence in favour of the asymptotic safety scenario has been accumulated in this framework.
There have also appeared several studies that no longer rely on the single field approximation and
nevertheless find additional evidence for a non-perturbative fixed point 
\cite{Manrique:2010am,Manrique:2009uh,Manrique:2010mq}, one of them working with fully fledged gravity \cite{Manrique:2010am}.

However, as mentioned in
the introduction, for one version of the analogous flow equation \eqref{equ:FRGE} in the infinite-dimensional $f(R)$ truncation
in asymptotic safety using the single field approximation, surprising results 
resembling the unexpected results of sec. \ref{sec:missingvacua} have been
found \cite{DietzMorris:2013-2}.
It is important to note that the choice of cutoff in the $f(R)$ approximation
in asymptotic safety is a complicated mixture of the separately relatively simple choices of field dependence
we have considered in \eqref{equ:cutoffh} with \eqref{cutoff1} or even \eqref{cutoff2}, \cf \cite{Benedetti:2012,Codello:2008,Machado:2007}.
Whereas only odd eigenoperators of all fixed points
except the Wilson-Fisher fixed point become redundant in the present case when $\al<0$,
the whole eigenspace collapses to a point in the $f(R)$ truncation of asymptotic safety.
However, the conceptual similarity between the two problems and their successful resolution in the scalar field case
in sec. \ref{sec:sWI-LPA} surely suggests 
 that the corresponding version of the sWI for gravity can help in resolving
such issues in asymptotic safety, leading to a more reliable description
of the fixed point and eigenoperator structure in the asymptotic safety scenario, 
even though the background field dependence of the effective average action
in gravity is intrinsic and therefore the direct elimination of the background field by a change of variable 
cannot be expected.

\vfill\break

\bibliographystyle{hunsrt}
\bibliography{refs}

\end{document}